\documentclass[twocolumn,tighten]{aastex631}
\hypersetup{linkcolor=red,citecolor=blue,filecolor=cyan,urlcolor=magenta}
\usepackage{amsmath}
\usepackage{natbib}
\usepackage{todonotes}
\usepackage{CJK}
\usepackage{xspace}
\usepackage[shortlabels]{enumitem}

\newcommand{\code}[1]{\texttt{#1}\xspace}
\newcommand{\Gaia}{{\it Gaia}\xspace}
\newcommand{\Latte}{\textit{Latte}\xspace}
\newcommand{\Fire}{\textsc{FIRE}\xspace}
\newcommand{\Ananke}{\code{Ananke}}
\newcommand{\agama}{\texttt{AGAMA}}
\newcommand{\degree}{$^{\circ}$}
\newcommand{\kmsec}{\mbox{km~s$^{\rm -1}$}}

\newcommand{\msun}{\mbox{$M_{\odot}$}}

\newcommand{\RN}[1]{%
  \textup{\uppercase\expandafter{\romannumeral#1}}%
}

\newcommand{\MnorE}{$2.15^{+1.22}_{-1.33}\times10^{11}$~\msun}
\newcommand{\rscaE}{$7.51^{+2.51}_{-3.07}$~kpc}
\newcommand{\slopE}{$1.46^{+0.75}_{-0.48}$}
\newcommand{\MvirE}{$1.50^{+0.31}_{-0.19}\times10^{11}$~\msun}
\newcommand{\rvirE}{$112.14^{+7.20}_{-4.92}$~kpc}
\newcommand{\concE}{$12.01^{+0.97}_{-0.79}$}
\newcommand{\rhoSE}{$0.491^{+0.049}_{-0.041}$~GeV\,cm$^{-3}$}
\newcommand{\chisE}{$0.04$}

\newcommand{\Mnorg}{$3.73^{+1.40}_{-1.03}\times10^{11}$~\msun}
\newcommand{\rscag}{$10.42^{+5.97}_{-4.18}$~kpc}
\newcommand{\slopg}{$0.90^{+0.34}_{-0.52}$}
\newcommand{\Mvirg}{$7.32^{+1.98}_{-1.53}\times10^{11}$~\msun}
\newcommand{\rvirg}{$190.13^{+15.79}_{-14.34}$~kpc}
\newcommand{\concg}{$17.17^{+3.44}_{-2.63}$}
\newcommand{\rhoSg}{$0.374^{+0.021}_{-0.022}$~GeV\,cm$^{-3}$}
\newcommand{\chisg}{$0.10$}

\shorttitle{Rotation curve on FIRE}
\shortauthors{Ou et al.} 

\begin{document}
\begin{CJK*}{UTF8}{gbsn}

\title{Decoding the Galactic Twirl: The Downfall of Milky Way-mass Galaxies Rotation Curves in the FIRE Simulations}

\author[0000-0002-4669-9967]{Xiaowei Ou (欧筱葳)} 
\affiliation{%
Department of Physics and MIT Kavli Institute for Astrophysics and Space Research, \\ 
Massachusetts Institute of Technology,\\
77 Massachusetts Avenue, Cambridge, MA 02139, USA}
\email{Email:\ xwou@mit.edu}

\author[0000-0003-2806-1414]{Lina~Necib}
\affiliation{Department of Physics and MIT Kavli Institute for Astrophysics and Space Research, \\
Massachusetts Institute of Technology,\\
77 Massachusetts Avenue, Cambridge, MA 02139, USA}
\affiliation{The NSF AI Institute for Artificial Intelligence and Fundamental Interactions, \\
Massachusetts Institute of Technology,\\
77 Massachusetts Avenue, Cambridge, MA 02139, USA}

\author[0000-0003-0603-8942]{Andrew~Wetzel}
\affiliation{Department of Physics \& Astronomy, University of California, Davis, CA 95616, USA}

\author[0000-0002-2139-7145]{Anna~Frebel}
\affiliation{%
Department of Physics and MIT Kavli Institute for Astrophysics and Space Research, \\ 
Massachusetts Institute of Technology,\\
77 Massachusetts Avenue, Cambridge, MA 02139, USA}

\author[0000-0001-6380-010X]{Jeremy~Bailin}
\affiliation{Department of Physics and Astronomy, University of Alabama, Box 870324, Tuscaloosa, AL 35487, USA}

\author[0000-0001-5636-3108]{Micah Oeur}
\affiliation{Department of Physics, University of California, Merced, 5200 Lake Road, Merced, CA 95343, USA}

\begin{abstract}

Recent measurements of the Milky Way rotation curve found a sharp decline at around $15$-$20$\,kpc from the center of the Galaxy, suggesting that the Galactic dark matter halo is much less massive than predicted by other dynamical tracers.
To address this tension, we study the validity of the assumptions made in calculating the Milky Way's rotation curve.
To do so, we apply Jeans' equation, the current standard approach of measuring rotation curves, to three cosmological zoom-in simulations of Milky Way-like galaxies from the \Fire-2 \Latte suite.
Using synthetic \Gaia\ surveys, we replicate the sample selection process and calculation employed in measuring the Milky Way rotation curve.
We examine four failure modes of this calculation and find that the measured curves deviate from the true curve by $5$-$20\%$ rather than below $5\%$, as estimated by previous works.
Interestingly, there is a large galaxy-to-galaxy variance, and different systematics dominate different galaxies.
We rederive the Milky Way's dark matter density profile with the rotation curve while incorporating systematics from the simulations.
The posterior distribution of the density profiles is consistent with a fiducial NFW profile when assuming a gNFW profile for dark matter. We find that the virial mass, \Mvirg, consistent with other probes of the Milky Way's mass.
However, we recommend that the field moves away from relying solely on the rotation curve when studying the dark matter profile, and adopts methods that incorporate additional probes and/or do not heavily depend on assumptions described in this study.

\end{abstract}


\section{Introduction}

Rotation curves of disc galaxies have long been fundamental tools for probing their mass distribution, offering a unique insight into the unseen dark matter that dominates their mass. 
By tracing the velocity of objects in near-circular orbits at various radii, these curves reveal the gravitational influence of both luminous matter and the much more massive dark matter halos that envelop galaxies \citep{rubin80}.

Within our own Milky Way, these dynamics provide a critical means to infer the total mass distribution, including dark matter, whose gravitational potential has a significant impact on the behavior of baryonic matter, i.e. stars and gas \citep{faber79}.
The local dark matter density in the vicinity of the Sun, derived from such rotation curves, is a key parameter in direct detection experiments \citep{Goodman85,Drukier86}, while the dark matter profile towards the Galactic center plays a pivotal role in indirect detection efforts \citep{chang08,cholis09,abramowski11,abdallah16,ackermann17}.
Further, the dark matter density profile at the inner Galaxy reflects a complex interplay of baryonic physics and dark matter physics at the galactic scale \citep[e.g.,][]{kazantzidis04,kaplinghat05,bullock05,chan15,tollet16,lazar20}.

The rotation velocity curve of the Milky Way has been measured with various tracers, with and without complete kinematics, depending on the galactocentric radii. 
The rotation curve at galactocentric radii within the solar radius can be derived from radio observation of the interstellar medium through the tangent-point method \citep{gunn79,fich89,levine08,sofue09}, requiring only three-dimensional kinematic information. 
At galactocentric radii outside of the solar radius, full six-dimensional kinematics are needed for any tracers used to constrain the rotation curve. 
Non-stellar tracers such as the thickness of the H\,\textsc{i} layer \citep{merrifield92}, spectrophotometric distances of H\,\textsc{ii} regions combined with radial velocities of associated molecular clouds \citep{fich89,brand93}, planetary nebulae \citep{schneider83}, and masers in high-mass star-forming regions \citep{reid14} are either rare or indirectly connected with the rotation curve through simplifying assumptions.
Stars, on the other hand, typically have relatively easily measurable distances, proper motions, and(or) line-of-sight velocities.
Namely, the stellar standard candles of classical cepheids \citep{pont97}, red clump giants \citep{bovy12,huang16}, RR Lyrae stars \citep{ablimit17,wegg19}, and blue horizontal branch stars \citep{xue09,kafle12} with radial velocity measurements are viable tracers of the underlying dark matter profile. 
Further, the emergence of astrometric surveys like \Gaia\ has significantly broadened the scope of high-precision parallax and proper motion measurements for general non-standard candle stars \citep{gaia16,gaia18,gaia21}. 
When paired with line-of-sight velocity data from large-scale spectroscopic surveys such as APOGEE \citep{majewski17} and LAMOST \citep{zhao12}, this allows for the mapping of the Galaxy's rotation curve out to large galactocentric distances \citep{eilers19,wang22,zhou22,ou24}.

Recent galactic rotation curve measurements, however, have found inconsistent results of the dark matter distribution compared with the results from other probes in the Milky Way.
Studies (\citet{wang22,zhou22,ou24}) found a faster decline in the rotation curve, suggesting a significantly less massive dark matter halo for the Milky Way.
\citet{jiao23} pointed out that this can be explained by a Keplerian decline, signaling the dark matter halo terminates exponentially at $R\sim20$\,kpc rather than extends to $\sim200$\,kpc following a power law decline.
These interpretations are in tension with the direct mass constraints from other probes at the outer dark matter halo, such as globular clusters \citep{eadie19,correamagnus22}, dwarf satellites \citep{callingham19,vasiliev21}, and stellar streams \citep{koposov23}. 

To address these inconsistencies, recent studies have focused on examining the accuracy of the curve measurements and the procedure of inferring dark matter profiles from the measured curve.
\citet{koop24} examined the second-order kinematic moments as a function of positions in the Milky Way and the cosmological simulations.
They concluded that merger events could induce axis asymmetry in the second-order kinematic moments and contribute to the systematic uncertainties of $\sim15\%$ in the final curve.
Observationally, studies using both dwarfs and giants have also identified clear evidence of axis asymmetry in the galactic thin disk \citep{wang22,tang24}.
The Jeans equation, which relies on axisymmetry, thus, is expected to be less reliable at the outer disk.
In addition, \citet{oman24} pointed out that the curve measurements at different galactic radii should not be treated as independent measurements, and the correlation, when properly accounted for, would significantly weaken the constraining power of the overall measurement.
Applying this reasoning to the \citet{ou24} results, the rotation curve would thus not suggest a statistically significant lower Milky Way mass despite the apparent decline at large radii.
More generally, such a behavior is not unique to the Milky Way rotation curve from stellar tracers. 
\citet{sands24}, using the \Fire-2 simulations, found the rotation curves of dwarf galaxies from H\,\textsc{i} tracers can deviate from the true rotation curves by up to $50\%$, resulting in an artificial ``diversity problem.''

In light of the observed axis asymmetries in the Milky Way stellar population and their potential impact on the rotation curve measurements, we conduct an end-to-end robustness test of the Jeans procedure to quantify the expected contribution from different sources of systematic uncertainties.
This is the first study testing the robustness of the Jeans procedure with stellar tracers, covering observational effects from the stellar population selection function, the model systematics from the inaccurate or omitted asymmetric drift correction terms, and the effects of axis-asymmetries and dynamical disequilibrium in the gravitational potential of the galaxies.
Using three Milky Way-mass galaxies from the \Latte suite of \Fire-2 simulations \citep{wetzel16, hopkins18} and corresponding synthetic surveys, we performed the Jeans procedure for rotation curve measurements identical to those used in recent Milky Way studies \citep{jiao23,ou24}.
We find that, even in galaxies with a quiet merger history, the Jeans equation and its assumptions are subject to significant systematic biases and uncertainties of order $5$-$20\%$.

We divide the Sections as following: in Section~\ref{sec:sim_survey}, we briefly describe the simulations and synthetic surveys used in this study. 
We provide an overview of the Jeans procedure and the four main sources of systematic uncertainties in Section~\ref{sec:methods}.
We also outline the specific methodologies we adopt to quantify the contribution from each source.
Results of the tests are shown in Section~\ref{sec:results} followed by interpretation and impact on our understanding of the Milky Way measurements in Section~\ref{sec:discussion}.

\section{Simulation and Synthetic Survey}
\label{sec:sim_survey}

In this work, we seek to compare the recovered rotation curve calculated through the Jeans equation (Sec.~\ref{sec:methods}), with the \textit{true} rotation curve obtained from simulations. To study any possible discrepancies arising from galaxy-to-galaxy variance, along with the effect of the selection of the stellar population, we use three Milky Way-mass simulated galaxies and the \Gaia synthetic surveys that correspond to these galaxies.  

\subsection{Cosmological Simulations of the Milky Way}
\label{sec:sims}

We use three zoom-in simulations of Milky Way-mass galaxies (m12i, m12f, and m12m) from the \Latte suite \cite{2016ApJ...827L..23W} of the \Fire-2 simulations \citep{hopkins15,hopkins18}, which are publicly available \citep{2023ApJS..265...44W}.
The simulations self-consistently model baryonic processes, such as star formation and metal enrichment. 
Additionally, they integrate the effects of galaxy formation in a cosmological context, incorporating a realistic history of mergers and accretion events \citep[for a detailed discussion of the simulations, see][]{2016ApJ...827L..23W,hopkins18,sanderson18,2023ApJS..265...44W}.

These galaxies are chosen because they reasonably resemble a realistic Milky Way-mass galaxy \citep{wetzel16,sanderson18}.
Specifically, studies have demonstrated that these simulated galaxies have galactic bar morphology \citep{debattista19,ansar25}, stellar thin and thick disk morphology \citep{ma17}, gas kinematics \citep{el-badry18,mccluskey24}, etc., that are broadly similar to the Milky Way.
It is impossible, however, for any simulated galaxy to fully reproduce all properties of the Galaxy.
Therefore, we interpret the findings from the simulations in the context of the Milky Way, keeping in mind these potential differences.

\subsection{\Gaia DR3 Synthetic Surveys}
\label{sec:syn_survey}

Apart from the simulations, we also utilize the synthetic \Gaia DR3 surveys of these three galaxies \citep{nguyen24}.
Briefly, these synthetic surveys are generated using \Ananke \footnote{\url{http://ananke.hub.yt/}} \citep{sanderson20}, based on the publicly available \textsc{PyAnanke} code base\footnote{\url{https://github.com/athob/py-ananke}} \citep{2024JOSS....9.6234T}, which samples a population of individual stars from simulated star particles, assigning them realistic physical properties and applying a selection function and error model to produce mock \Gaia observations. 
We refer readers to \citet{sanderson20} and \citet{nguyen24} for details regarding the synthetic surveys. 

These synthetic surveys provide a realistic test ground to examine observational constraints-induced biases in stellar kinematic analyses using \Gaia\ data.
Specifically, we use these synthetic surveys to test the effect of the \Gaia\ selection function on the stellar sample for measuring rotation curves (see Section~\ref{sec:failure_modes}).

\subsection{Spectroscopic Survey(s) Selection Function}
\label{sec:apogee_sf}

In addition to \Gaia, some recent Milky Way rotation curve measurements also used stellar spectra from surveys such as APOGEE \citep{majewski17} and LAMOST \citep{cui12} to improve the distance measurements and provide more accurate radial velocities \citep{wang22,ou24}. 
While possible, modeling the selection function of ground base telescopes with fiber-fed spectroscopy is more challenging \citep[see e.g.][]{bovy16b} and is currently unavailable for the \Fire simulations.
Nonetheless, for this study, we partially factor in the selection function of ground-based spectroscopy on bright red giant stars by selecting stars with low surface gravity ($\log{g}<2.2$) to replicate the selection made in \citet{ou24}. 
We expect the full spectroscopic survey selection function, if fully recovered, to worsen any existing biases we find in this study as a realistic spectroscopic survey selection function are generally more limited.
The results in this study, thus, should be treated as a lower limit for potential biases introduced by the sample selection.

\section{Rotation curve calculation}
\label{sec:methods}

In this Section, we review the standard procedure of the rotation curve calculation from stellar tracers (Section~\ref{sec:jeans_procedure}) and clarify the nomenclature between rotation and circular velocity curves (Section~\ref{sec:rot_vs_circ}). 
We list the common assumptions and simplifications adopted in recent curve measurements, and describe their potential failure modes (Section~\ref{sec:failure_modes}).
A description of different sets of rotation curves we calculated using the simulations and synthetic surveys for quantifying the failure modes is presented as well (Section~\ref{sec:curves_sets}).

\subsection{Jeans procedure}
\label{sec:jeans_procedure}

The full form of the collisionless Boltzmann equation in cylindrical coordinates $(R,\varphi,z)$ is written as
\begin{equation}
\begin{aligned}
    \frac{\partial f }{\partial t} + p_R \frac{\partial f}{\partial R} + &\frac{p_\varphi}{R^2} \frac{\partial f}{\partial \varphi} + p_z \frac{\partial f}{\partial z} - \left( \frac{\partial \Phi}{\partial R} - \frac{p^2_\varphi}{R^3} \right) \frac{\partial f}{\partial p_R} \\ 
    - &\frac{\partial \Phi}{\partial \varphi} \frac{\partial f}{\partial p_\varphi} - \frac{\partial \Phi}{\partial z} \frac{\partial f}{\partial p_z} = 0,  
\label{eq:boltzmann_full}
\end{aligned}
\end{equation}
where $f$ is the 6D phase space density distribution of the tracer population with potential time ($t$) dependence, $p_{R,\varphi,z}$ are the momenta in cylindrical coordinates,\footnote{We follow the standard notations $(r,\phi,\theta)$ and $(R,\phi,z)$ for spherical and cylindrical coordinates, respectively, in the galactocentric frame, unless otherwise noted.} and $\Phi$ is the confining potential.

Assuming that the Milky Way's gravitational potential is axisymmetric (i.e., no $\varphi$ dependence) and in dynamical equilibrium (i.e., no $t$ dependence), the equation simplifies to 
\begin{equation}
    p_R \frac{\partial f}{\partial R} + p_z \frac{\partial f}{\partial z} - \left( \frac{\partial \Phi}{\partial R} - \frac{p^2_\varphi}{R^3} \right) \frac{\partial f}{\partial p_R} - \frac{\partial \Phi}{\partial z} \frac{\partial f}{\partial p_z} = 0.
\label{eq:boltzmann}
\end{equation}
Multiplying the equation by $p_R$ and integrating over the momenta $p_{R,\varphi,z}$, we obtain the Jeans equation in cylindrical coordinates \citep{jeans15,binney08}, which links the moments of the velocity distribution and the density of a collective of stars to the gravitational potential via
\begin{equation}
    \frac{\partial \nu \langle v^2_R\rangle}{\partial R} + \frac{\partial\nu\langle v_R v_z\rangle}{\partial z} + \nu\left(\frac{\langle v^2_R\rangle - \langle v^2_{\varphi}\rangle}{R} + \frac{\partial\Phi}{\partial R} \right) = 0,  
\end{equation}
where $\nu$ is the density distribution of the tracer population. 
The rotation velocity ($v_{\rm{rot}}$), which is directly linked to the confining potential, can then be calculated as
\begin{equation}
     v^2_{\rm{rot}} = R\frac{\partial \Phi}{\partial R} = \langle v^2_{\varphi}\rangle - \langle v^2_R\rangle - \frac{R}{\nu} \left( \frac{\partial \nu \langle v^2_R\rangle}{\partial R} + \frac{\partial\nu\langle v_R v_z\rangle}{\partial z} \right).
    \label{eq:jeans}
\end{equation}

In practice, the kinematic moments ($\langle v^2_{\varphi,R}\rangle$ and $\langle v_{R} v_z \rangle$) are directly calculated from observed stellar tracers.
Due to unknown selection functions, the tracer density profile ($\nu$) is often challenging to obtain from data and is thus assumed to be some functional form based on independent studies.
Likewise, depending on the tracer population distribution, the radial and vertical profiles ($\frac{\partial \langle v^2_{R}\rangle}{\partial R}$ and $\frac{\partial \langle v_R v_z\rangle}{\partial z}$) of the kinematic moments can be either determined from data or assumed to be some known profiles from the literature.

The common practice \citep[see, e.g.][]{eilers19,zhou22} is to omit the higher-order terms in the equation and assume exponential form for the underlying tracer population profile, simplifying the equation to 
\begin{equation}
    v^2_{\rm rot}(R) = \langle v^2_{\varphi}\rangle - \langle v^2_R\rangle\left(1 - \frac{R}{R_0} - \frac{2R}{R_1}\right), 
    \label{eq:vrot2}
\end{equation}
where $R_0$ is the scale length of the exponential tracer density ($\nu$) profile and $R_1$ is the scale length of the exponential radial velocity dispersion ($\sqrt{\langle v^2_R\rangle}$) profile. 
The $\sqrt{\langle v^2_R\rangle}$ term with its coefficients ($1 - \frac{R}{R_0} - \frac{2R}{R_1}$) corrects for the asymmetric drift in the stellar tracer population \citep{binney08}.

For the simulations considered in this study, we assume fixed $R_0$ values from \citet{sanderson18}, similar to what was done in the case of the Milky Way.
The latter, $R_1$, can be constrained from the stellar tracers, but we adopted a fixed $R_1$ value of $25$\,kpc, the best-fit value for the Milky Way from \citet{ou24}.
When allowing $R_1$ to change between each rotation curve calculation, we find that the value varies significantly across simulations (described in Section~\ref{sec:curves_sets}), with values occasionally being unphysically high ($>100$\,kpc) due to outliers in the stellar population.
Nonetheless, the values are consistently greater than $R_0$, making the term $R/R_0$ in Eq.~\ref{eq:vrot2} always dominate over the $2R/R_1$ term.
Thus, for simplicity and interpretability, we use a fixed $R_1$ value for all curves calculated in this study and note that the results remain unchanged if we leave $R_1$ as a free parameter and fit it for every curve calculation.

\subsection{Rotation vs. circular velocity curves}
\label{sec:rot_vs_circ}

In this study, we refer to the quantity calculated from Equation~\ref{eq:vrot2} as the rotation curve or rotation velocity, but note that this has also been referred to as the circular velocity curve in previous studies \citep[e.g.,][]{eilers19,ou24}.
In this study, the rotation curve denotes the theoretical velocity a tracer would be moving at if it is in perfectly circular orbit within the galactic disk.
The theoretical, true $v_{\rm rot} (R)$ is tied to the gravitational potential gradient (or gravitational acceleration) experienced by the tracer.
Equation~\ref{eq:jeans} calculates the observed $v_{\rm rot} (R)$, assuming axisymmetric and dynamical equilibrium, and thus does not necessarily reflect the true $v_{\rm rot} (R)$ or gravitational potential.

The nomenclature is not well-defined for the circular velocity curve, but typically is directly tied to the mass distribution of the galaxy,
\begin{equation}
    v_{\rm circ}(r) = \sqrt{\frac{GM(r)}{r}}, 
    \label{eq:vcirc2}
\end{equation}
where $G$ is the gravitational constant and $M(r)$ is the mass enclosed at a distance $r$ from the galaxy's center. 
The circular velocity ($v_{\rm circ}$) is a purely theoretical construct that reflects the mass distribution of the system, and the expression for $v_{\rm circ}$ in Equation~\ref{eq:vcirc2} does not rely on assumptions of symmetries or equilibrium in the mass distribution or potential so long as a center is defined.

We note that, despite the nomenclature confusion, all calculations regarding the Milky Way disk rotation/circular velocity curves mentioned in this and previous studies refer to the $v_{\rm rot}$.

\subsection{Failure Modes of the Jeans Equation}
\label{sec:failure_modes}

We identify and examine four potential sources of systematic uncertainties of the idealized Jeans equation method for calculating the rotation curve of a Milky Way-like galaxy.
These systematics manifest either as directional biases or additional non-directional uncertainties.
The four sources of systematic uncertainties are:
\begin{enumerate}[(i)]
    \item Biased stellar population from the survey selection function,
    \item Inaccurate/omitted asymmetric drift correction term,
    \item Dynamical disequilibrium due to recent mergers, 
    \item Axis asymmetry.
\end{enumerate}
We now discuss each in turn. 

\subsubsection{Biased Stellar Population}
\label{sec:mode_sf}

The first source of systematics, the biased stellar population from the survey selection function, arises from the sample of stars selected. This is due to either the selection function of the survey itself, or any additional cuts performed in the analysis. 

In rotation curve studies \citep[see e.g.][]{ou24}, the sample of stars is selected through (i) the \Gaia selection function, (ii) the APOGEE selection function, and (iii) additional cuts for the analysis. 

\textit{Gaia Selection Function:} Surveys such as \Gaia preferentially select brighter stars at larger distances as they are easier to observe. 
As a result, the more distant stars are biased to be more evolved giant stars, which typically have experienced more asymmetric drift due to dynamical interactions than main sequence stars of the same initial mass.

\textit{APOGEE Selection Function:} Ground-based spectroscopic surveys such as APOGEE also preferentially select red giant stars due to their high flux in the near-IR and relative insensitivity to galactic extinction.
Both the \Gaia\ and APOGEE selection functions are modeled by the synthetic surveys as described in Section~\ref{sec:syn_survey} and \ref{sec:apogee_sf}.

\textit{Additional Cuts:}
In addition to the survey selection function, recent studies of the Milky Way rotation curve also apply additional selection of thin disk stars to remove contamination from the halo and thick disk stars.
While halo stars experience the same gravitational potential as all other stellar tracers, their orbits are not rotationally supported and have dynamical timescales varying from hundreds of Myr to several Gyr.
Their kinematics cannot be modeled by the Jeans equation formalism.
On the other hand, thick disk stars are generally considered to be older and have different density and velocity dispersion profiles compared to thin disk stars, requiring more complex modeling with a Jeans equation.
To fully simulate the selection of the recent rotation curve calculation studies, we thus apply the same sample cuts used in \citet{eilers19,ou24}. 
We find that the specific disk selections (described below) have minimal effects on the final measured curves in most cases examined here compared to other sources of systematic uncertainties.

More explicitly, we adopt the following additional cuts when deriving the rotation curves from the simulations and synthetic surveys to remove stars from the thick disk and stellar halo population and stars affected by the galactic bar. 
We first replicate the low $\alpha$-element abundances $[\alpha/\rm{Fe}]$ cut to select the thin disk stars.
We visually inspect the overall $[\alpha/\rm{Fe}]$ distributions in these \Fire-2 simulations and found systematic offsets from the Milky Way distribution, likely due to offsets in the assumed stellar nucleosynthesis models \citep{carrillo23}.
We thus place the cut at $[\alpha/\rm{Fe}]<0.27$ based on visual inspection to remove thick-disk stars. 
To remove contamination from the halo, we select stars with (1) low velocities perpendicular to the galactic plane ($|v_z| < 100$\,\kmsec) and (2) low heights above/below the galactic plane ($|z| < 1$\,kpc or $|z|/R < \tan{(\pi/30)}$). 
We remove stars potentially affected by the non-axisymmetric potential near the galactic bars of these three simulated galaxies \citep{ansar25} with the cut $R > 6$\,kpc. 
Lastly, we limit our sample to within a wedge of $60$\,\degree around the assumed solar position for a given synthetic survey (see Table~1 of \citet{nguyen24}).
Combining the selection cuts for the disk stars on top of the survey selection function discussed above, we examine if the more biased stellar population at the outer disk contributes to any significant systematic offset.

\subsubsection{Asymmetric Drift Correction}
\label{sec:mode_hoc}
The second source of systematics, the inaccurate/omitted asymmetric drift correction term in the Jeans equation, arises because these terms depend on the density and velocity dispersion profiles of the tracer population, which are often not well constrained by the observed data \citep{koop24}. 
Therefore, studies typically omit the higher-order terms in Equation~\ref{eq:jeans} and assume an exponential form for the underlying tracer population profile.

While the magnitude of the omitted higher-order terms ($\frac{\partial\nu\langle v_R v_z\rangle}{\partial z}$) is expected to be small, it is nonetheless expected to be more significant at the outer disk, where disk flaring may contribute more significant drift in the vertical direction \citep{bovy16}. 
Additionally, for first-order terms ($\frac{\partial \nu \langle v^2_R\rangle}{\partial R}$) currently included, the Jeans equation requires prior knowledge of the stellar tracer distribution in phase space, particularly the density and radial velocity dispersion profiles. 
Constraining the true profiles of the observed tracer stars is possible yet often challenging due to the unknown selection function of the survey taking the data. 
The typical practice, which we replicate in this study, is to assume an exponential form with a fixed scale radius based on independent studies. Below, we study the effect of such an assumption.

\subsubsection{Axis Asymmetry}
\label{sec:mode_aa}
The third source of systematics, axis asymmetry, arises due to the potential being non-axisymmetric. 
The Jeans equation (Equation~\ref{eq:jeans}) assumes axisymmetry, which may not be true even for galaxies with a quiet accretion history.
In the Milky Way, merger events such as the Large Magellanic Cloud (LMC) and Sagittarius (Sgr) \citep{ibata94} can introduce distortion in the disk shape, resulting in disk warping, further deviating the stellar disk from axis symmetry \citep[see e.g.][]{laporte18}.

\subsubsection{Dynamical Disequilibrium}
\label{sec:mode_dd}
The last source of systematics, dynamical disequilibrium, arises because the stars may not trace the underlying potential, especially if the host galaxy has experienced a recent major merger. 
If there had been major merger events within the free-fall timescale of the Milky Way, the stars would not have had enough time to adjust to the new potential, and thus would have led to inaccurate estimates for the rotation curve.
Such time variance may theoretically be captured in the time-dependent term in the full collisionless Boltzmann equation. 
The information needed to perform such a calculation is unfortunately out-of-reach observationally for the Milky Way and computationally expensive even with simulation data. 

\subsection{Quantifying the Systematics}
\label{sec:curves_sets}

We calculate three sets of rotation curves (Table~\ref{tab:curves}) using different methodologies on different samples to examine the presence and significance of the four sources of systematic uncertainties discussed in Section~\ref{sec:failure_modes}. 
Specifically, we calculate the following rotation curves:

\begin{enumerate}
    \item \textbf{\Ananke\ curves from Jeans equation:} Jeans equation calculation (Equation~\ref{eq:vrot2}) with stars from the \Ananke\ synthetic surveys at present day ($z=0$). These curves are set to represent the observable ones that could be calculated for the Milky Way.
    \item \textbf{\Fire-2 curves from Jeans equation:} Jeans equation calculation with disk star particles from the simulations at varying redshifts and azimuths. These curves are set to represent the result of the Jeans equations, but without being affected by any survey selection functions.
    \item \textbf{\Fire-2 curves from \agama\ fitted potential:} Curves derived from taking derivatives of the \agama\ fitted potential with all particles from the simulations at varying redshifts. These curves are set to represent the \textit{true} potential, albeit assuming axisymmetry. 
\end{enumerate}

For all curves from sets (1) and (2), we adopt the Jeans equation to calculate the rotation curve from the stellar kinematics measured in cylindrical galactocentric coordinates, as described in Section~\ref{sec:failure_modes}. 
The galactocentric coordinates are determined using the code \code{GizmoAnalysis}.\footnote{\url{https://bitbucket.org/awetzel/gizmo_analysis}}
The Jeans equations assume dynamical equilibrium and axisymmetry while omitting higher-order asymmetric drift correction terms.
The curves from set (3) are directly derived from the fitted potential and its gradient, which is assumed to be axisymmetric when fitting. 
The curve in this case, thus, is not dependent on the tracer used but only deviates from the true curve if there is axis asymmetry in the gravitational potential.\footnote{We briefly discussed the intrinsic axis asymmetry of the gravitational potential in Section~\ref{sec:res_aa}.}
For this study, we mainly discuss axis asymmetry in the context of whether the distribution of the stars used for the curve measurements is azimuthally angle-dependent.

In terms of sample selection, for both sets (1) and (2), the star (particles) used for deriving any given curve is limited to a 60\,\degree range in azimuth and binned in $R$ at $0.5$\,kpc intervals, with larger bins adopted at $R\gtrsim20$\,kpc to account for the reducing number of available star tracers.
The set (3) curves, however, are derived without azimuthal cuts and do not require binning in $R$ as the fitted potential gives an analytical expression for the true rotation curve.
We apply the same disk selection described in \cite{ou24}, and discussed in Sec.~\ref{sec:mode_sf}. 
Therefore, comparisons between sets (1) and (2) mainly constrains the observational effects in rotation curve calculations, whereas comparisons between sets (2) and (3) focus on the accuracy of the Jeans equation (Eq.~\ref{eq:vrot2}).
Additionally, curves in set (2) are unique because they are derived with samples of star particles at varying azimuthal angles and time steps, whereas set (1) is only derived at $z=0$ and set (3) is derived using particles at all angles.

Table~\ref{tab:curves} briefly summarizes the different sets of curves and corresponding assumptions.

\begin{deluxetable*}{cccc}
\caption{
Summary of curves derived in this study for each of the three galaxies, and the associated failure modes and assumptions.
\label{tab:curves}
}
\tablehead{
\colhead{} & \colhead{\textbf{\Ananke\ curves}} & \colhead{\textbf{\Fire-2 curves}} & \colhead{\textbf{\Fire-2 curves}} \\
\colhead{} & \colhead{\textbf{from Jeans equation}} & \colhead{\textbf{from Jeans equation}} & \colhead{\textbf{from \agama\ fitted potential}}
}
\startdata
Selection function-biased tracer population & Affected & Unaffected & Unaffected \\
\hline
Higher order asymmetric drift correction & Affected & Affected & Unaffected \\
Inaccurate tracer profiles &  &  &  \\
\hline
Dynamical disequilibrium & Affected & Affected & Unaffected \\
\hline
Axis asymmetry  & Affected & Affected & Unaffected\tablenotemark{$\ast$} 
\enddata
\tablenotetext{\ast}{The fit is performed on all simulation particles but principally assumes axis symmetry. See texts for details.}
\end{deluxetable*}

\section{Results}
\label{sec:results}

\begin{figure}
    \centering
    \includegraphics[width=0.95\linewidth]{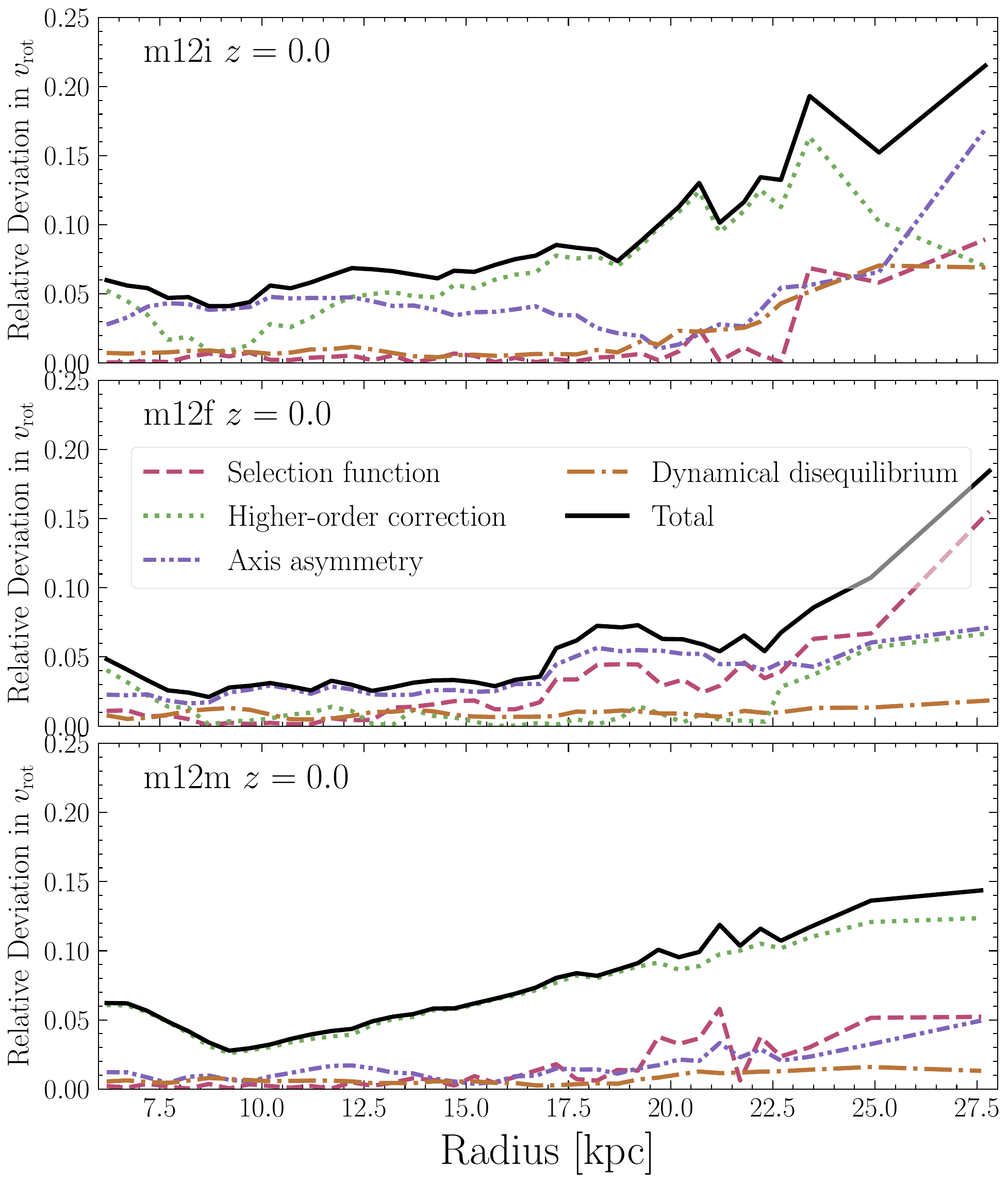}
    \caption{
    Summary of all sources of systematic uncertainties in rotation curve measurements analyzed in this study. 
    All uncertainties are relative to the curve derived from \agama\ (set 3).
    From top to bottom, the three panels show the results for different simulated galaxies at $z=0$ (snapshot 600): m12i, m12f, and m12m, respectively.
    The systematic uncertainties from dynamical disequilibrium (dashed-dotted line) are moderate ($<5\%$) in most cases (except for cases when there are recent merger events; see Section~\ref{sec:res_dd}).
    The systematic uncertainties from the selection function (dashed line) are moderate for m12i and m12m, but more significant ($>5\%$) for m12f.
    The systematic uncertainties from axis asymmetry (dashed-double-dotted line) and higher-order correction (dotted line) are significant in most cases.
    }
    \label{fig:sys_sum_rel}
\end{figure}

We present the results for all systematic uncertainties in Figure~\ref{fig:sys_sum_rel} for the three simulated galaxies. 
There are notable differences between the magnitude of systematic uncertainties for the three galaxies.
For m12i, the systematic uncertainties are dominated by higher-order correction and axis asymmetry, with selection function and dynamical disequilibrium effects only becoming significant at the outer disk.
For m12f, the axis asymmetry plays a more important role at the inner disk, with the sample selection growing steadily and becoming dominant at the outer disk.
Lastly, for m12m, the total systematic uncertainty is dominated by the higher-order correction, with the rest three contributing only moderately.
Regardless of the relative contribution from different systematic sources in all three galaxies, the total uncertainties are typically $>5\%$ and reach $\sim20\%$ at the outer disk.
This is even more significant than the previous estimate on the Milky Way rotation curve systematic uncertainties, which range between $2$-$15\%$ \citep{ou24}.

\subsection{Effect of the Biased Stellar Population}
\label{sec:res_sf}

\begin{figure*}
    \centering
    \includegraphics[width=0.32\linewidth]{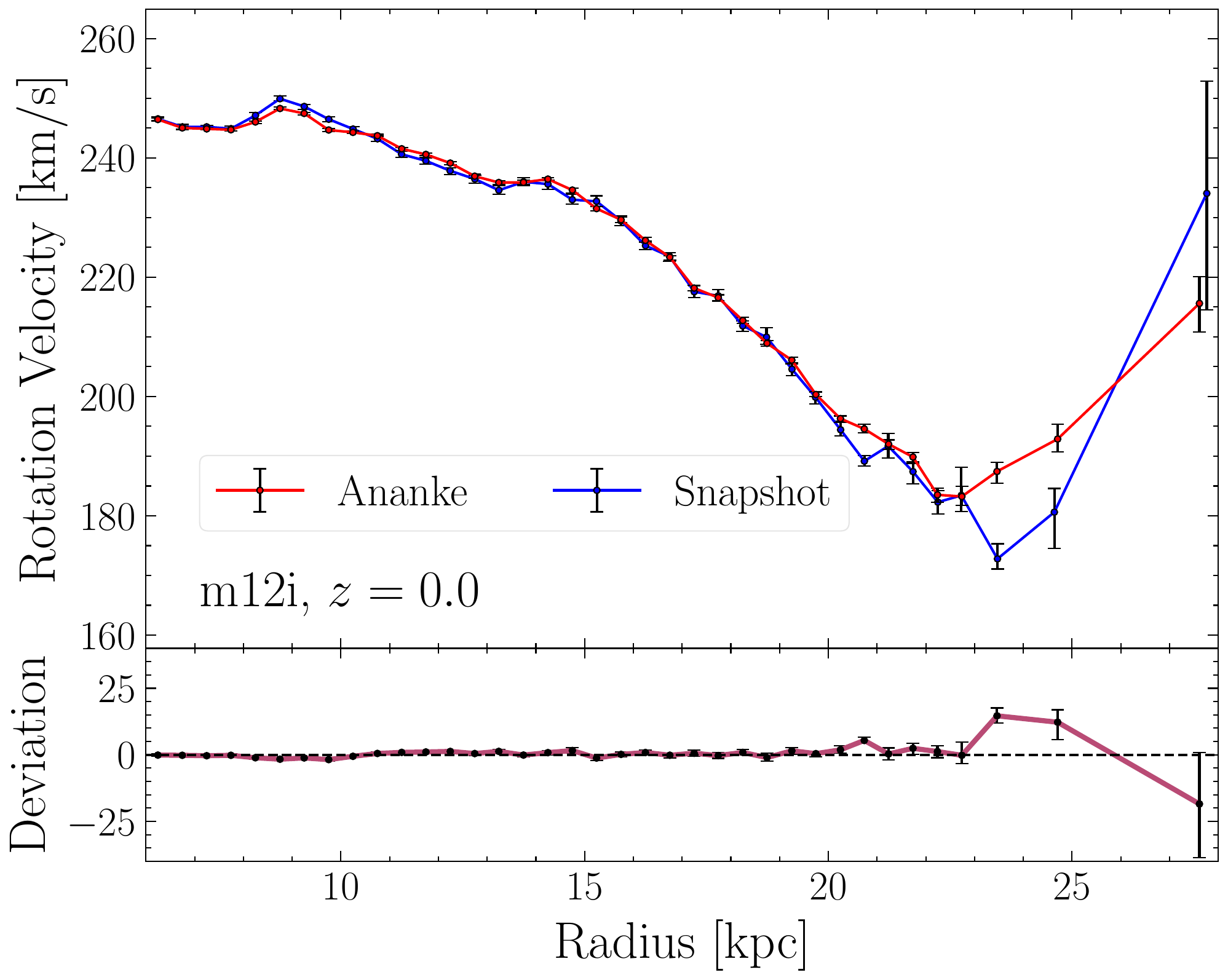} 
    \includegraphics[width=0.32\linewidth]{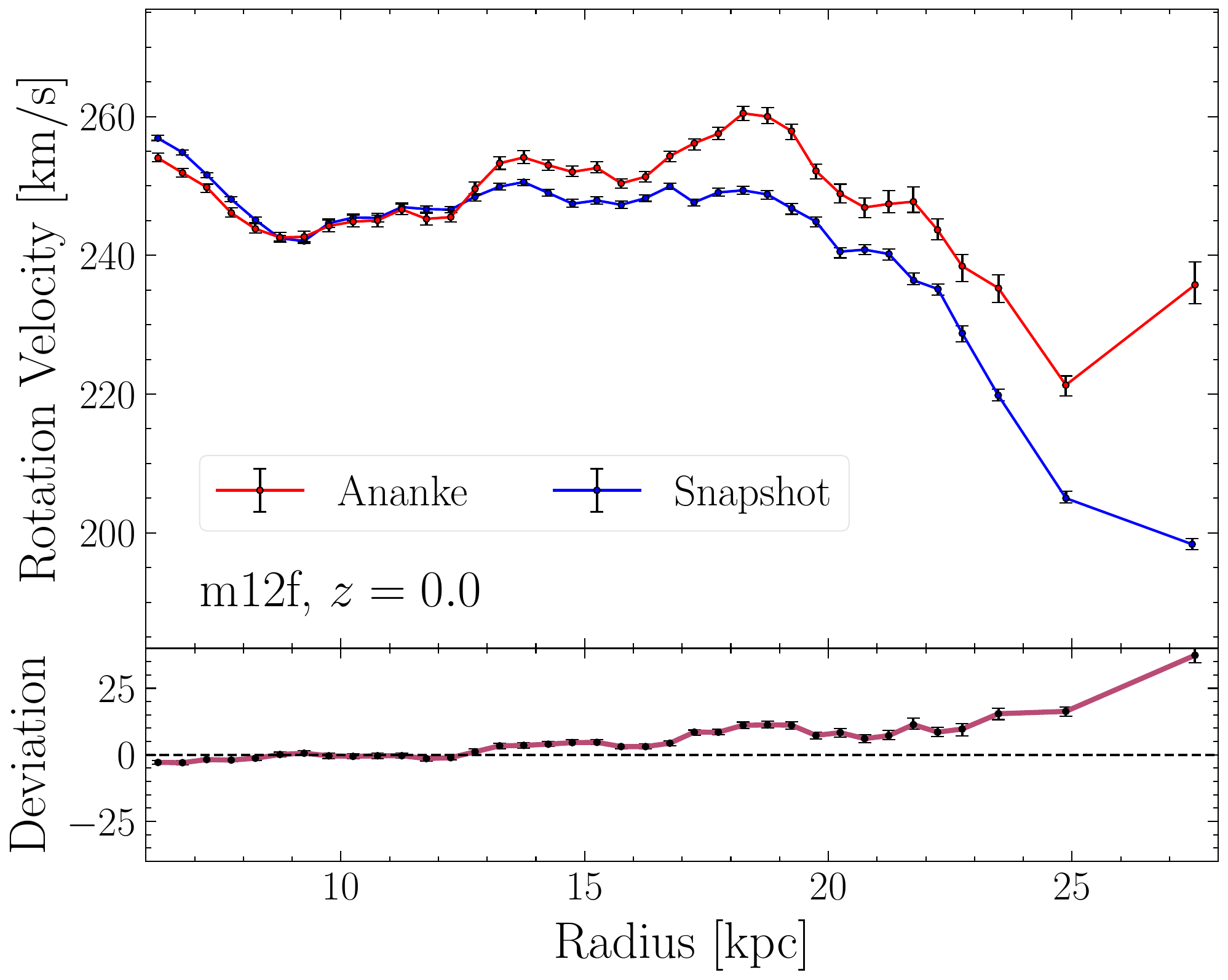} 
    \includegraphics[width=0.32\linewidth]{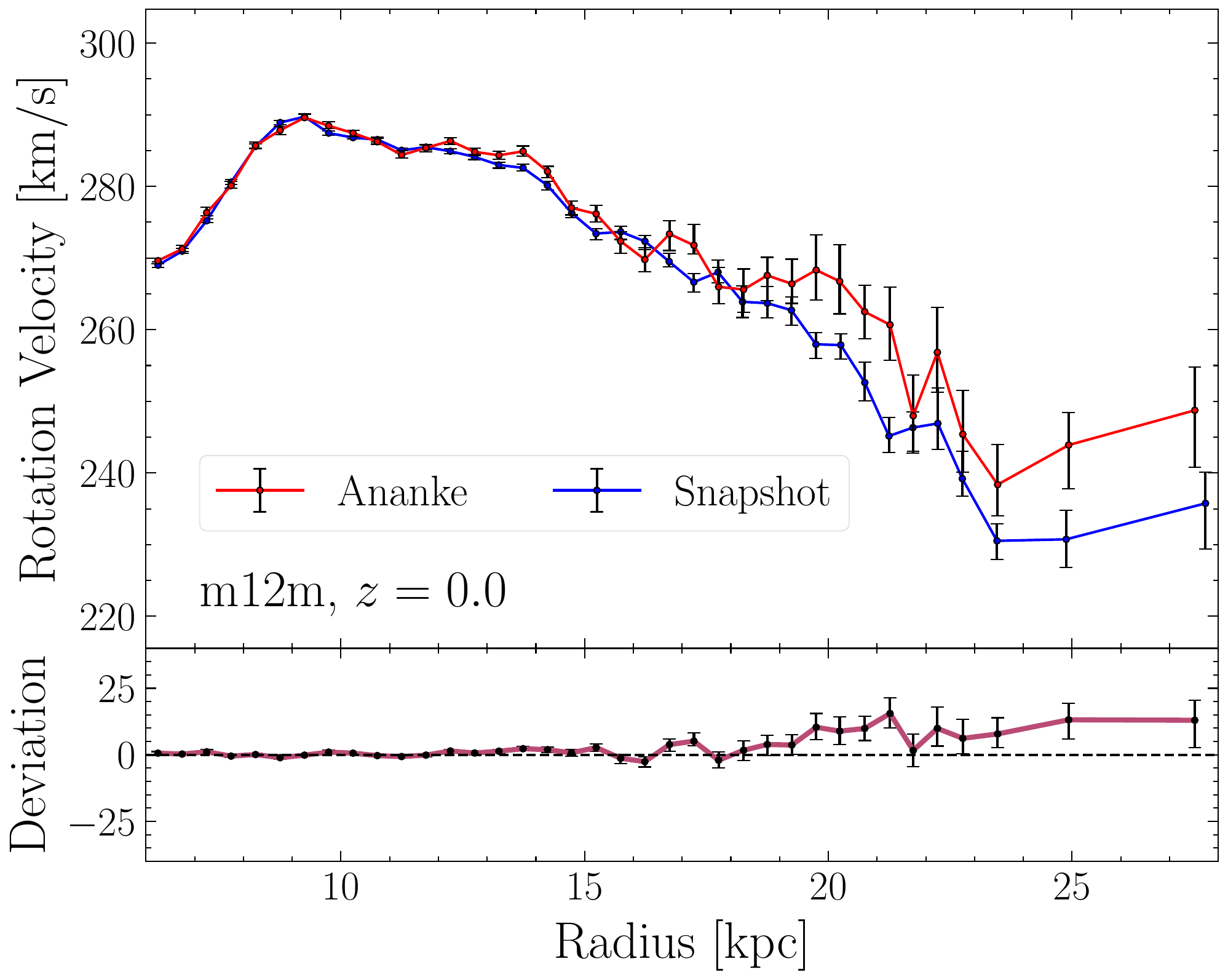}
    \caption{
    Comparison between the rotation curves from the synthetic \Gaia\ survey (set (1); red dots) and the rotation curves from the star particles in the simulation at $z=0$ (set (2); blue dots).
    The error bars show the corresponding statistical uncertainties in the curves.
    From left to right, the three plots show the results for different simulated galaxies at $z\sim0$ (snapshot 600): m12i, m12f, and m12m, respectively.
    The top panel of each plot shows the derived rotation curve as a function of radius from the galactic center.
    The bottom panel shows the absolute difference between the \Ananke\ and snapshot curves. 
    The error bars in the differences are calculated by adding the statistical uncertainties from the two curves in quadrature.
    The line style of the bottom panel is matched to Figure~\ref{fig:sys_sum_rel}. 
    }
    \label{fig:sys_sf_rel}
\end{figure*}

To disentangle the effect of the survey selection function used in the analysis, we compare the curves from sets (1) and (2), specifically those measured at $z=0$ and the same azimuth. 
These two sets differ mainly in the presence of the biased sample selection due to the survey characteristics.
To emphasize the effects of the survey selection function, we still apply the additional cuts described in Section~\ref{sec:mode_sf} as those selections are arbitrarily defined and change between studies.
We note that these additional cuts have a negligible effect on the measured rotation curve compared to the survey selection functions and other systematic sources examined.

The rotation curve derived from the \Ananke\ sample shows minimal to moderate deviation from the curve derived from snapshot star particles.
For m12f and m12m, the \Ananke\ curves are overestimated compared to the snapshot results at large $R\gtrsim17$\,kpc, as shown in Figure~\ref{fig:sys_sf_rel}.
The deviation is less significant for m12i and appears only for $R\gtrsim22$\,kpc.
We discuss below the likely cause of the overestimated \Ananke\ curves at the outer disk. 

As pointed out in Section~\ref{sec:failure_modes}, the overestimated \Ananke\ results can be caused by inaccurate scale radii for the assumed and observed stellar tracer profiles due to different selection function at the inner vs. outer disk.
Specifically, while the currently included asymmetric drift correction term ($\frac{\partial \nu \langle v^2_R\rangle}{\partial R}$) in Eq.~\ref{eq:jeans}, accounts for the drift in the radial direction, the accuracy of such a correction relies on the assumed density and radial velocity dispersion profiles.
The scale radii adopted in our \Ananke\ measurements are primarily determined by stars at the inner disk due to the larger sample size.
However, due to the sample selection, the \Ananke\ stellar sample preferentially selects more evolved RGB stars at the outer disk, which have intrinsically more different density and radial velocity dispersion profiles than the full stellar population considered in the \Fire-2 snapshot measurements. 
A recent study with eleven galaxies from \Fire-2 simulations \citep{mccluskey24} has shown that the ratios of $v_{\phi}$ to total velocity dispersion, i.e., how rotationally supported the stars are, decrease with stellar ages.
Furthermore, stars older than 8\,Gyr gained most of their velocity dispersion through dynamical heating after star formation. 
In other words, these RGB stars have experienced more asymmetric drift due to their interaction with other stars.
The asymmetric drift results in a higher average $v_R$ for these stars at the outer disk.
At the same time, their density and radial velocity dispersion profiles are driven outwards, with longer tails towards greater $R$.

As a result, while $\langle v^2_R\rangle$ was driven higher due to the biased sample, the assumed scale radii, $R_0$ and $R_1$, were not adjusted accordingly to account for the increasing dominance of RGB stars in the tracer population.
The net result is that the \Ananke\ measurements overestimate the asymmetric drift correction and produce higher values in comparison with the \Fire-2 measurements that use all star particles.
Consequently, growth in a disagreement as a function of $R$ is expected as the selection bias to RGB stars becomes more significant at the outer disk.

Enforcing an additional age cut on the \Ananke\ sample to reduce the number of the evolved RGB stars can mitigate this effect and bring the curve to agree with the \Fire-2 snapshot curve.
The ideal solution, however, is to divide and characterize the tracer profiles with different ages and perform the measurement separately.
Such analysis is challenging for the Milky Way as the stellar age estimates often have significant uncertainties.
The limited sample size at the outer disk also prevents statistically reliable modeling of the profiles as a function of stellar age. 

The overall bias extent of this is of order 10-20 km/s, meaning $\sim10\%$ relative error, for m12i and m12m. 
For m12f, the bias reaches $\sim40$\,\kmsec at the outer disk, dominating the total uncertainties.

\subsection{Effect of the Asymmetric Drift Correction}
\label{sec:res_hoc}

\begin{figure*}
    \centering
    \includegraphics[width=0.32\linewidth]{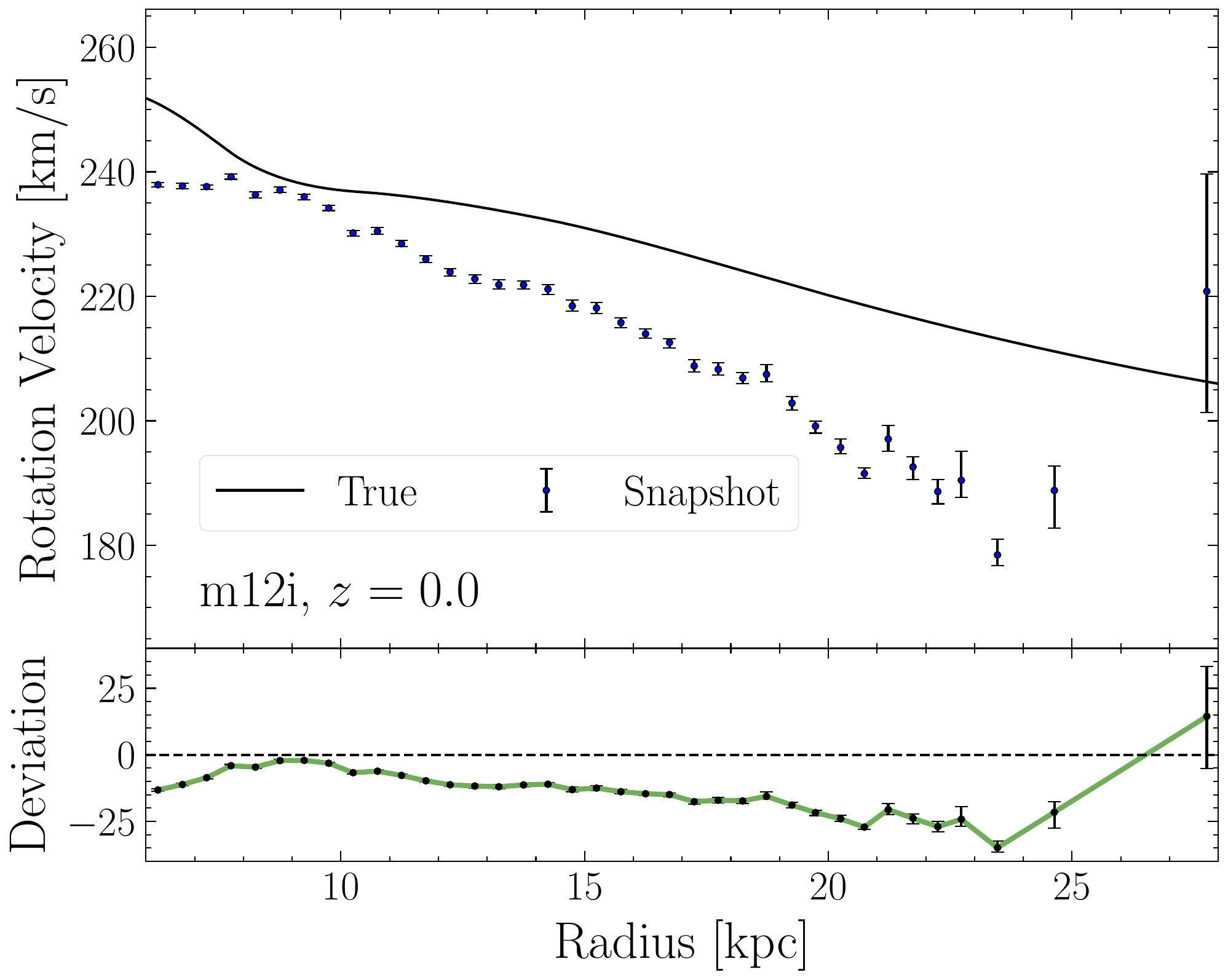} 
    \includegraphics[width=0.32\linewidth]{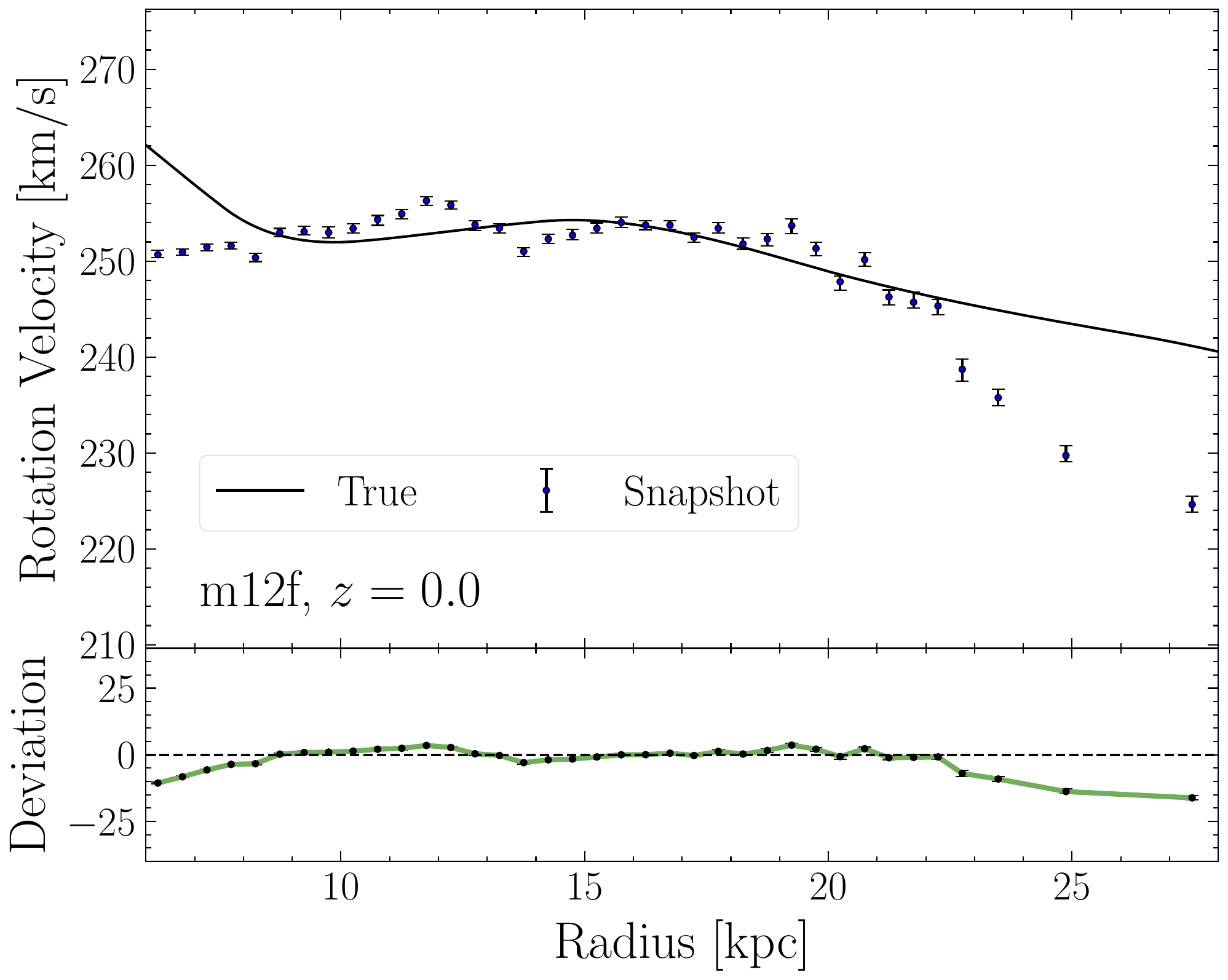} 
    \includegraphics[width=0.32\linewidth]{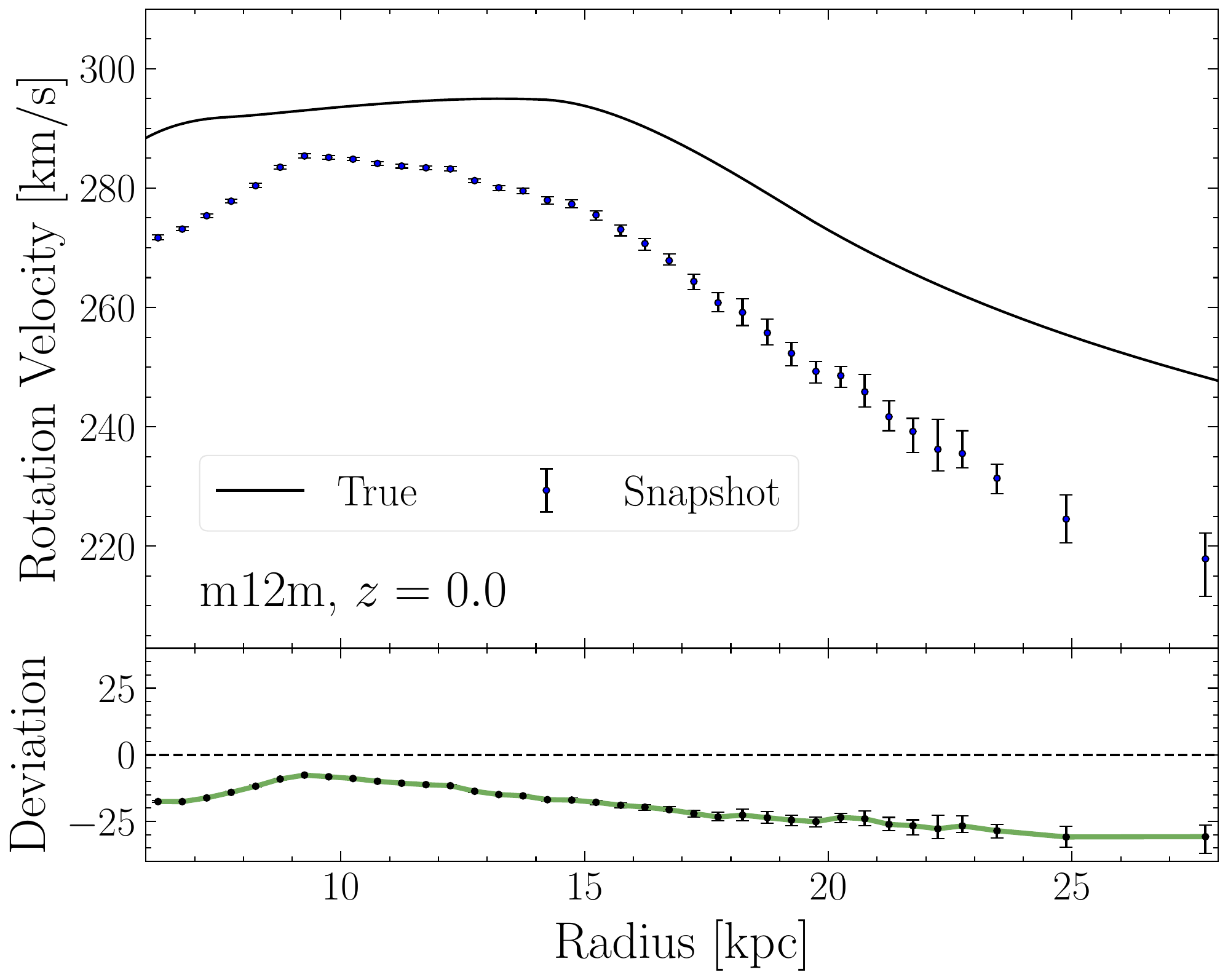}
    \caption{
    Similar to Figure~\ref{fig:sys_sf_rel} but for comparison between the simulation snapshot curves (set (2); blue dots) and the \agama\ curves directly derived from the potential at $z=0$ (set (3); black curve).
    From left to right, the three plots show the results for different simulated galaxies at $z\sim0$ (snapshot 600): m12i, m12f, and m12m, respectively.
    For each galaxy, the top panel shows the derived rotation curve as a function of radius from the galactic center.
    The bottom panel shows the difference between the snapshot and \agama\ curves. 
    }
    \label{fig:sys_hoc_rel}
\end{figure*}

For the omitted higher-order asymmetric drift correction term and any potential inaccuracies in the tracer profiles, we compare curves between sets (2) and (3).
We note that the deviation between sets (2) and (3) can also be attributed to the dynamical disequilibrium and axis asymmetry.
Thus, we take the azimuth angle-averaged curves from set (2) for the comparison to eliminate the effect of axis asymmetry.
Additionally, we choose the curves in set (2) that do not show significant time evolution across the snapshots within a characteristic dynamical timescale.

The rotation curve derived from the simulation snapshot particles following the Jeans equation shows a significant deviation from the curve derived from \agama\ fitted potential (Figure~\ref{fig:sys_hoc_rel}).
All three galaxies show that the \agama\ results are higher than the snapshot results at almost all $R$, and the deviation increases at larger $R$.
For m12f, the deviation is insignificant until $R\sim22$\,kpc, despite increasing to $\sim20$\,\kmsec\ at the outer disk, similar to the cases of m12i and m12m.

The sign and growing magnitude of the disagreement are expected as the asymmetric drift is most significant at the outer disk.
In particular, the disk flaring moves kinetic energy into the vertical velocity dispersion, which is only captured by the higher-order terms \citep{binney08}. 
Ignoring the higher-order correction terms thus naturally results in underestimated curve measurements.

The differences could also be attributed to inaccuracies in modeling the stellar tracer density and dispersion profiles. 
Note that this is related to, but a broader issue than, what was discussed in Section~\ref{sec:res_sf}.
Exponential profiles may not accurately describe the stellar tracer profile at the outer disk due to disk flaring and warping.
These issues are intrinsic to the disk stars' distributions before the selection functions are factored in.

In Figure~\ref{fig:sys_hoc_rel}, the deviations between the true and snapshot rotation curves are least significant at the inner disk, where we expect minimal asymmetric drift in the vertical direction and the exponential profile approximation to be accurate.
Nonetheless, a recent study \citep{mccluskey24} has shown non-negligible vertical velocity dispersions ($\sigma_{v_z}$) across the star formation history of these \Fire-2 galaxies.
Future follow-up studies will examine the $R$ dependence of the velocity dispersions of stars and quantify the degree of disk flaring and warping in \Fire-2 galaxies, which will explain the different behaviors of the deviations as a function of $R$ between the three galaxies. 

Similar studies have been conducted in the Milky Way to examine the effects of inaccurate asymmetric drift correction on rotation curve measurements. 
\citet{koop24} found evidence of the Milky Way tracer density profile and velocity dispersion profile showing deviations from the typically assumed exponential form.
They also identified non-negligible vertical dispersion in the disk stars and estimated their contribution to the higher-order drift correction.
They estimated that these factors could result in the Jeans equation calculated rotation curves deviating from the true rotation curves by $\sim10$-$15\%$.
This is consistent with our results showing an overall scatter size of order 10-40 km/s, corresponding to $\sim10$-$20\%$ relative error.

\subsection{Effect of Axis asymmetry}
\label{sec:res_aa}

\begin{figure*}
    \centering
    \includegraphics[width=0.32\linewidth]{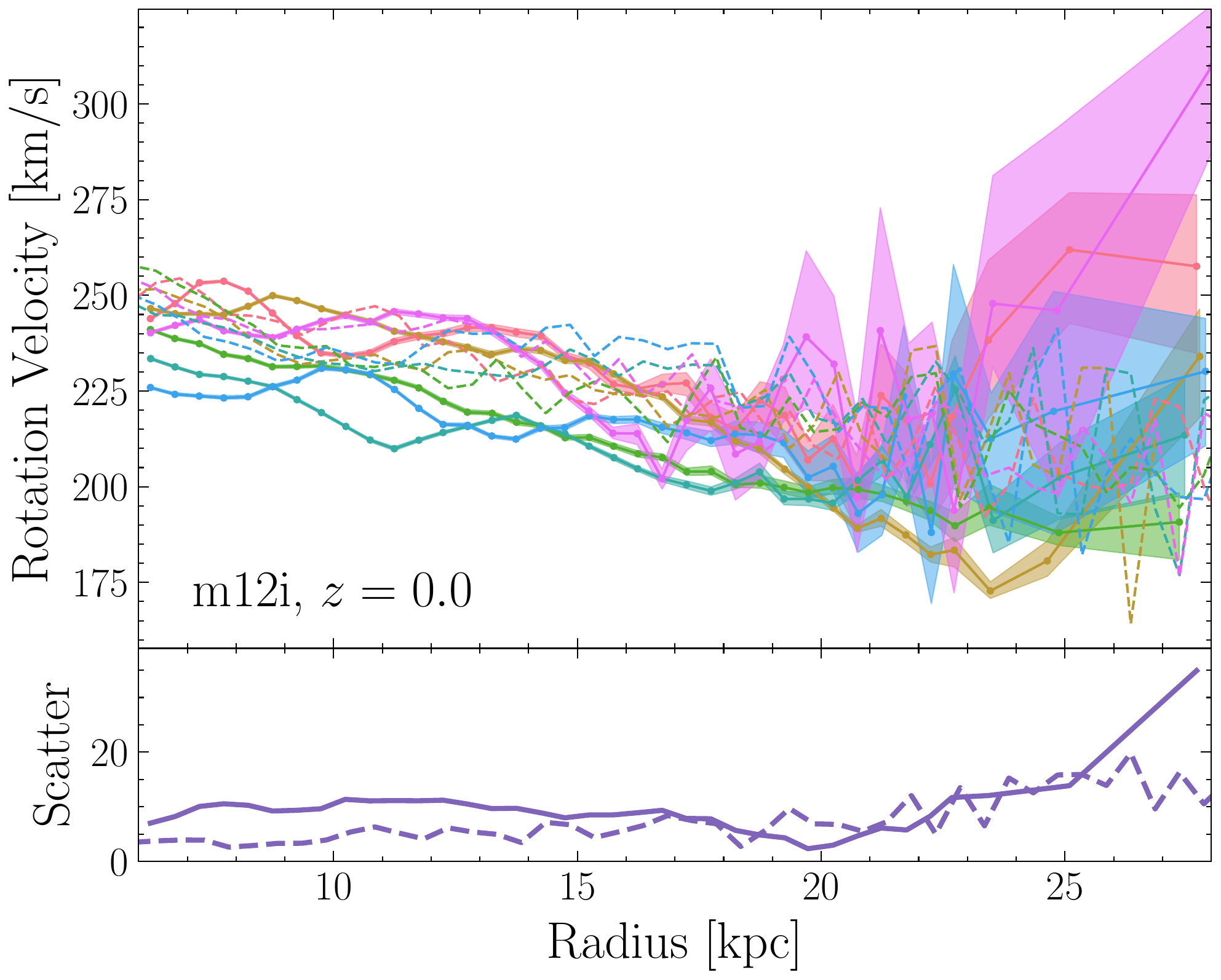} 
    \includegraphics[width=0.32\linewidth]{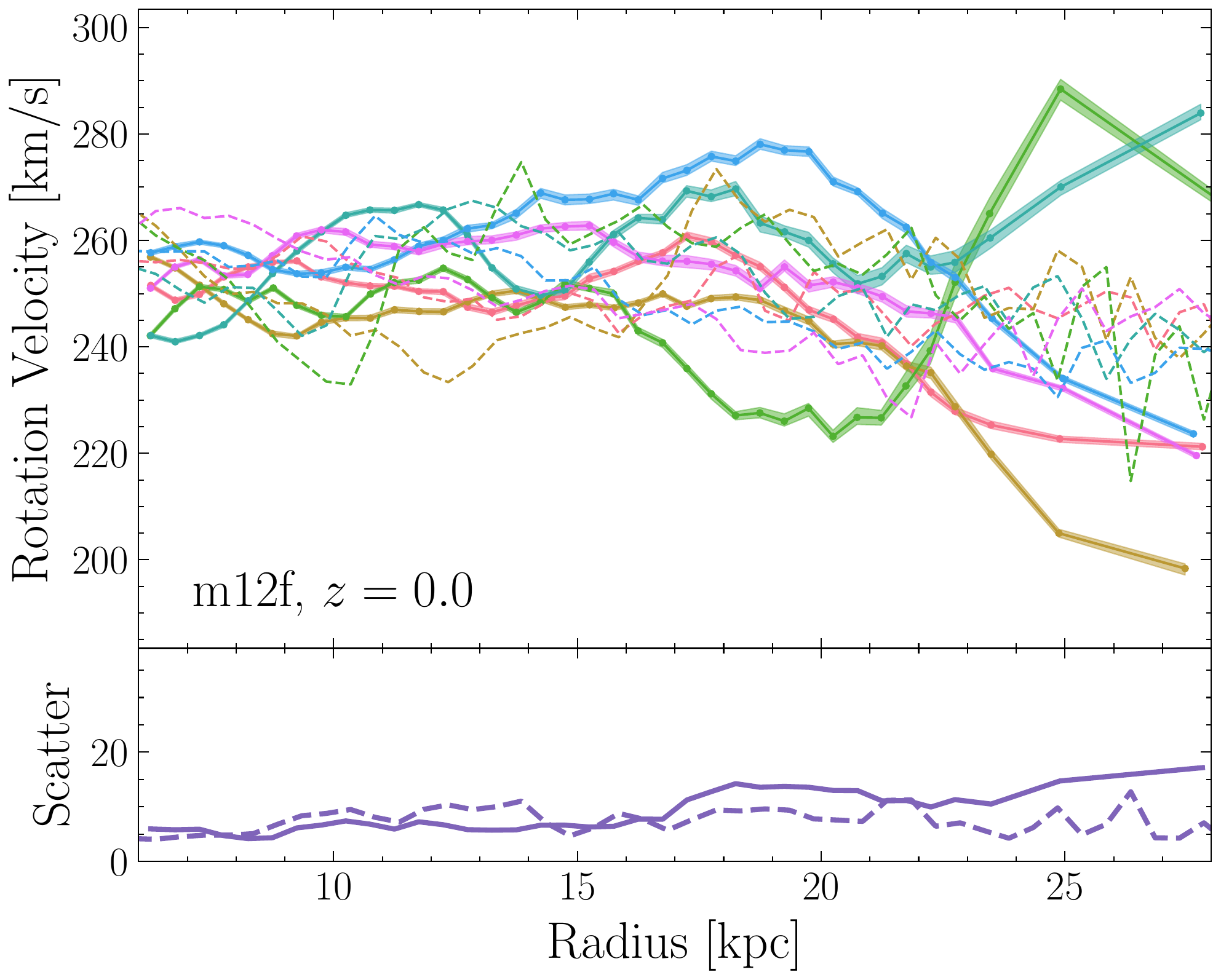} 
    \includegraphics[width=0.32\linewidth]{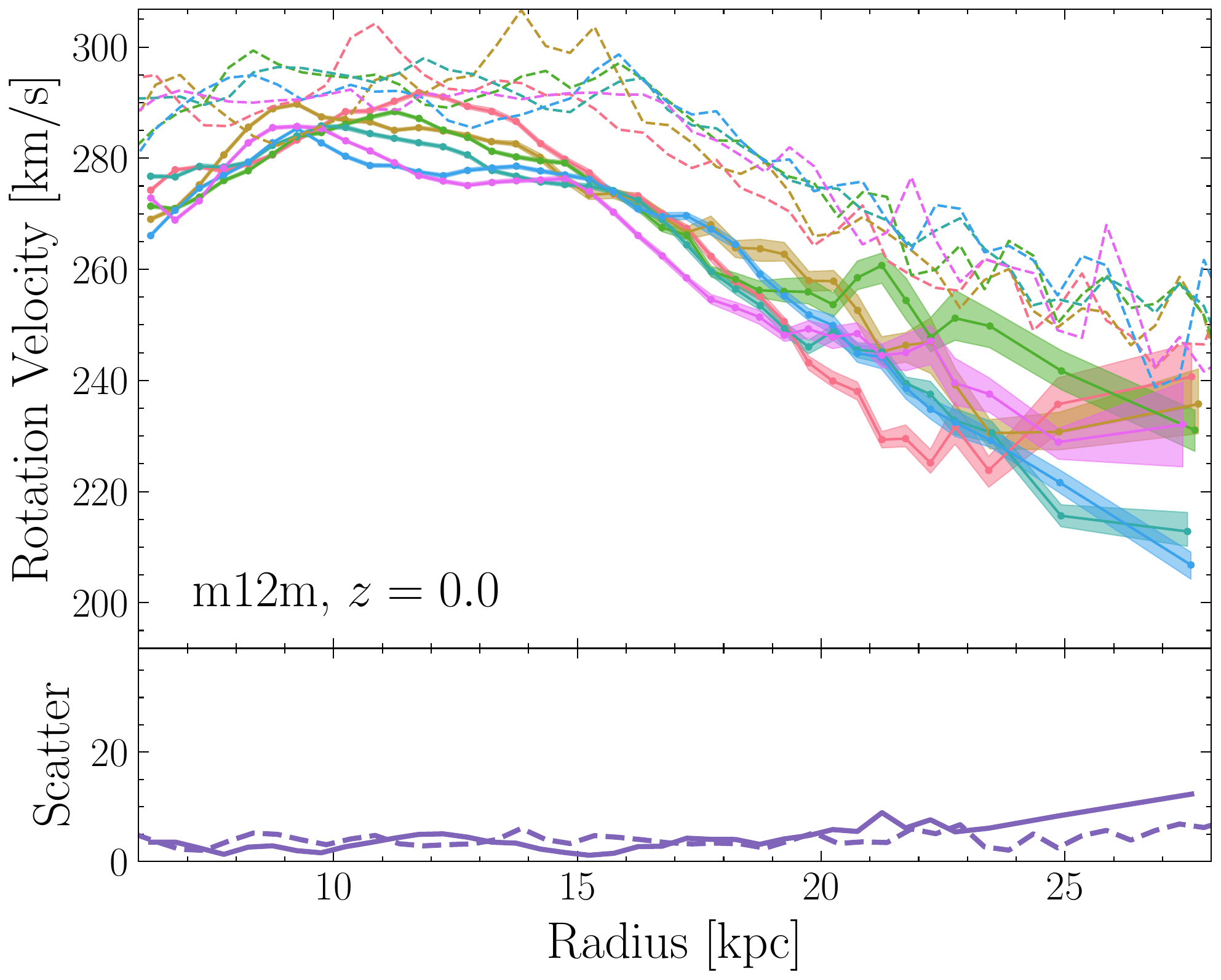}
    \caption{
    Comparison between the simulation snapshot curves (set (2)) six non-overlapping 60\,\degree\ wedges at different azimuthal angles (solid lines) at $z=0$.
    The color bands show the corresponding statistical uncertainties in the curves.
    The dashed lines represent the rotation curves calculated directly from the snapshots recorded gravitational potential (see text for details).
    The scatter in the dashed line, while present, is not as large as the scatter in the solid lines, indicating asymmetries both in the potential and the stellar tracers.
    From left to right, the three plots show the results for different simulated galaxies at $z\sim0$ (snapshot 600): m12i, m12f, and m12m, respectively.
    For each galaxy, the top panel shows the derived rotation curve as a function of radius from the galactic center based on stars in six different wedges.
    The solid and dashed lines in the bottom panel show the scatter in the solid and dashed rotation curves, respectively. 
    }
    \label{fig:sys_aa_rel}
\end{figure*}

The rotation curves show a significant scatter even at the same snapshots, as shown in Figure~\ref{fig:sys_aa_rel}. 
The curves shown are calculated from star particles in six non-overlapping 60\,\degree\ wedges at different azimuthal angles in set (2). 
The shaded region of each rotation curve represents the statistical uncertainties associated with the curve.
The statistical uncertainties are extremely small, especially in the inner galaxy, thus emphasizing that the scatter cannot be explained by poor statistics (i.e., lack of stars) in a particular azimuthal range of the galaxies.

The scatter between the different curves (shown in the bottom panel of Fig.~\ref{fig:sys_aa_rel}) suggests that the curve we derive is angle-dependent, thus breaking the axis-symmetry assumption.
We further examine if the angular variation in the curves is due to the intrinsic variation in the potential.
To do so, we examine the simulation snapshots' recorded gravitational potential (up to some arbitrary scaling) at the location of each star particle. 
We are \textit{not} using the \agama-fitted axisymmetric potential (curves (3)) in this case as such potential assumes an axisymmetry (as we noted in Sec.~\ref{sec:curves_sets}). 
Briefly, we calculate the average potential experienced by star particles within the azimuthal angle and galactocentric distance bins (same as those used for computing the sets (1) and (2) curves).
With the average potentials, we perform numerical derivatives to compute the rotation curves at each 60\,\degree azimuthal bins, shown as dashed lines in Figure~\ref{fig:sys_aa_rel}. 

\begin{figure}
    \centering
    \includegraphics[width=0.95\linewidth]{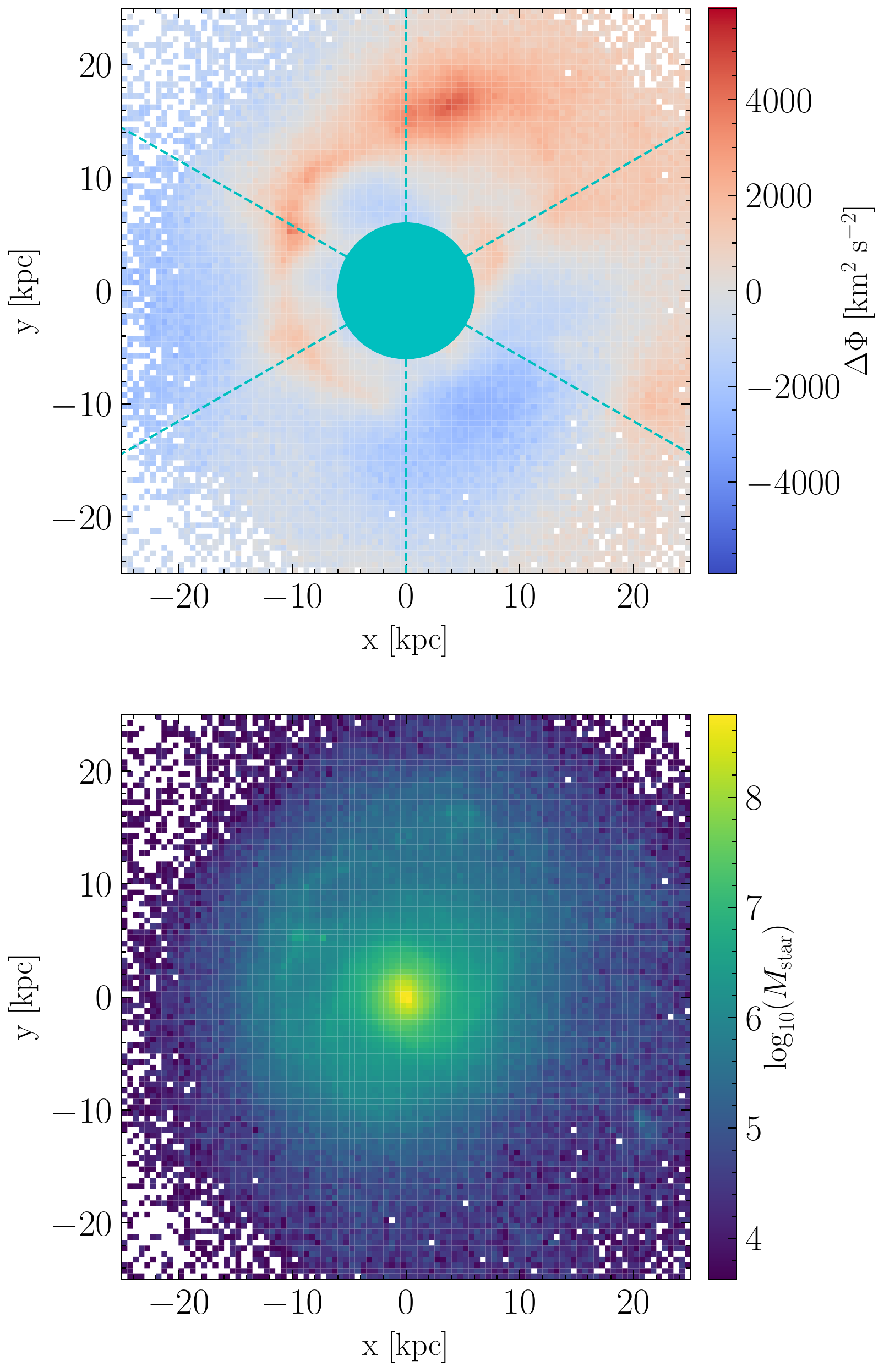} 
    \caption{
    Comparison between the axis asymmetric potential experienced by star particles (top) and the stellar mass (bottom) in the galactic plane for m12f.
    The \agama-fitted axisymmetric component is subtracted to highlight the azimuthal variation in potential.
    The cyan mask at the center of the top panel indicates the region not employed in the rotation curve calculation due to known non-axisymmetric potential driven by the galactic bar.
    The dashed lines mark the different azimuthal slices used to compute the rotation curves.
    The deviation in gravitational potential from complete axis symmetry in the top panel matches qualitatively with the stellar mass distribution in the stellar disk.
    }
    \label{fig:delta_pot_map}
\end{figure}

For all three galaxies, the variation in the potential appears small on the order of $1\%$,\footnote{Because the gravitational potential has arbitrary normalization, this fraction is estimated from the ratio between the fluctuation and the full range of potential of all star particles.} and the derived curves show variability on the order of 5-15 km/s.
The general offsets between the star tracer-derived curves (solid lines) and the potential-derived curves (dashed lines) in Figure~\ref{fig:sys_aa_rel} are driven by the asymmetric drift correction, as discussed in Section~\ref{sec:res_hoc}.
Here, we focus on comparing the scatters in these two sets of curves.

Overall, the scatter in the potential-derived curves is not negligible in size and qualitatively explains the axis-asymmetry we observe in our star tracer-derived rotation curves at $R\lesssim20$\,kpc.
For $R\gtrsim20$\,kpc, however, this intrinsic gravitational potential variation is insufficient to fully explain the axis-asymmetry we observe in the rotation curves. 
The additional scattering is partially driven by the lack of statistics at the outer disk, as shown by the more significant statistical uncertainties in the shaded areas in Figure~\ref{fig:sys_aa_rel} and sparse particles in Figure~\ref{fig:delta_pot_map} at $R>20$\,kpc.

Comparing galaxy to galaxy, we find that m12i has a greater scatter in the star tracer-derived rotation curves even at the inner galaxy ($R\lesssim15$\,kpc).
On the other hand, the scatters are comparable from $R\sim15$ to 25\,kpc for m12i, only with the star tracer-derived rotation curves scatter become greater than the potential-derived curves scatter at $R\gtrsim20$\,kpc.
For m12f and m12m, the scatters are comparable throughout the inner galaxy and only begin to diverge at $R\sim17$ and $R\sim23$\,kpc, respectively.
Therefore, there is no apparent universal relation between the axis asymmetries in the potential and the stellar tracer distribution. 
We thus only qualitatively characterize the axis asymmetries for the three galaxies and leave the physical interpretation of these galaxy-to-galaxy variances to future work.

We examine the axis asymmetry in the potential by subtracting the axis-symmetric \agama-fitted potential away from the total potential, focusing on the axis asymmetric fluctuation as shown in Figure~\ref{fig:delta_pot_map}.
We find that the fluctuation qualitatively matches the disk stellar mass distribution for all three galaxies.\footnote{See Appendix~\ref{sec:more_pot} for m12i and m12m.}
Such resemblance means that the rotation curves calculated from stars are sensitive to the non-axisymmetric gravitational potential components of the stars themselves at a $\sim1\%$ level.
The stellar component thus partially drives the total potential away from axis symmetry, perturbing the rotation curves calculated on the order of a few percent.
We note that this does not preclude other intrinsic azimuthal variations in the gravitational potential.
Triaxial dark matter halos, for example, create additional axis asymmetries that will be superimposed on top of the stellar contribution \citep[e.g.,][]{law09}. 
Previous work has also shown that the \Fire-2 galaxies gravitational potential are better represented by multipole models that utilize base function expansion \citep{arora22}.
Disentangling these contributions is beyond the scope of this study.

Overall, we see the gravitational potential itself and the stellar tracer distribution are not axisymmetric and exhibit a significant degree of angular variation.
The scattering in the derived rotation curves qualitatively traces the azimuthal variation in the potential.
The overall scatter size is 10-40\,\kmsec, meaning $\sim10$-$20\%$ relative error.

\subsection{Effect of Dynamical disequilibrium}
\label{sec:res_dd}

\begin{figure*}
    \centering
    \includegraphics[width=0.32\linewidth]{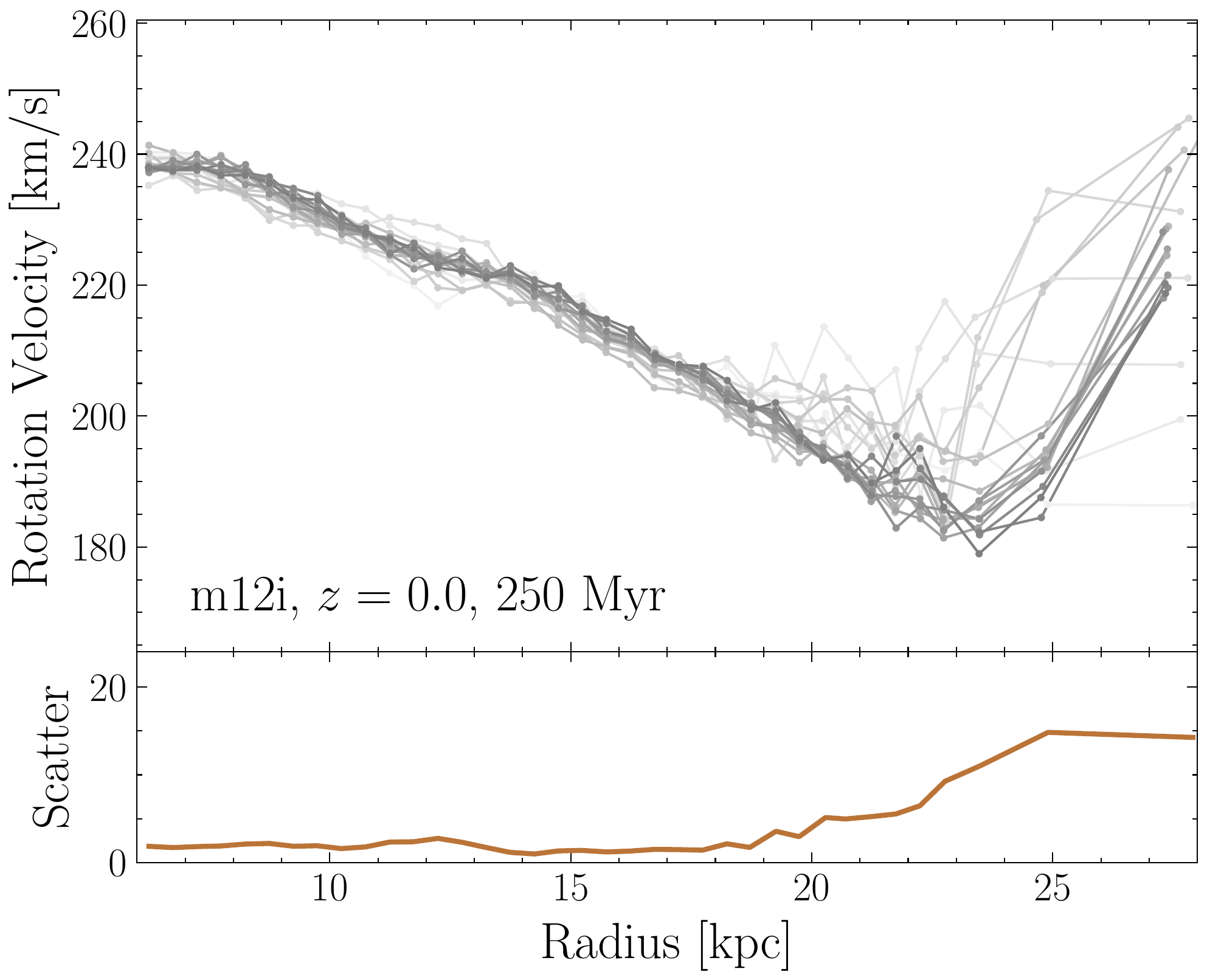} 
    \includegraphics[width=0.32\linewidth]{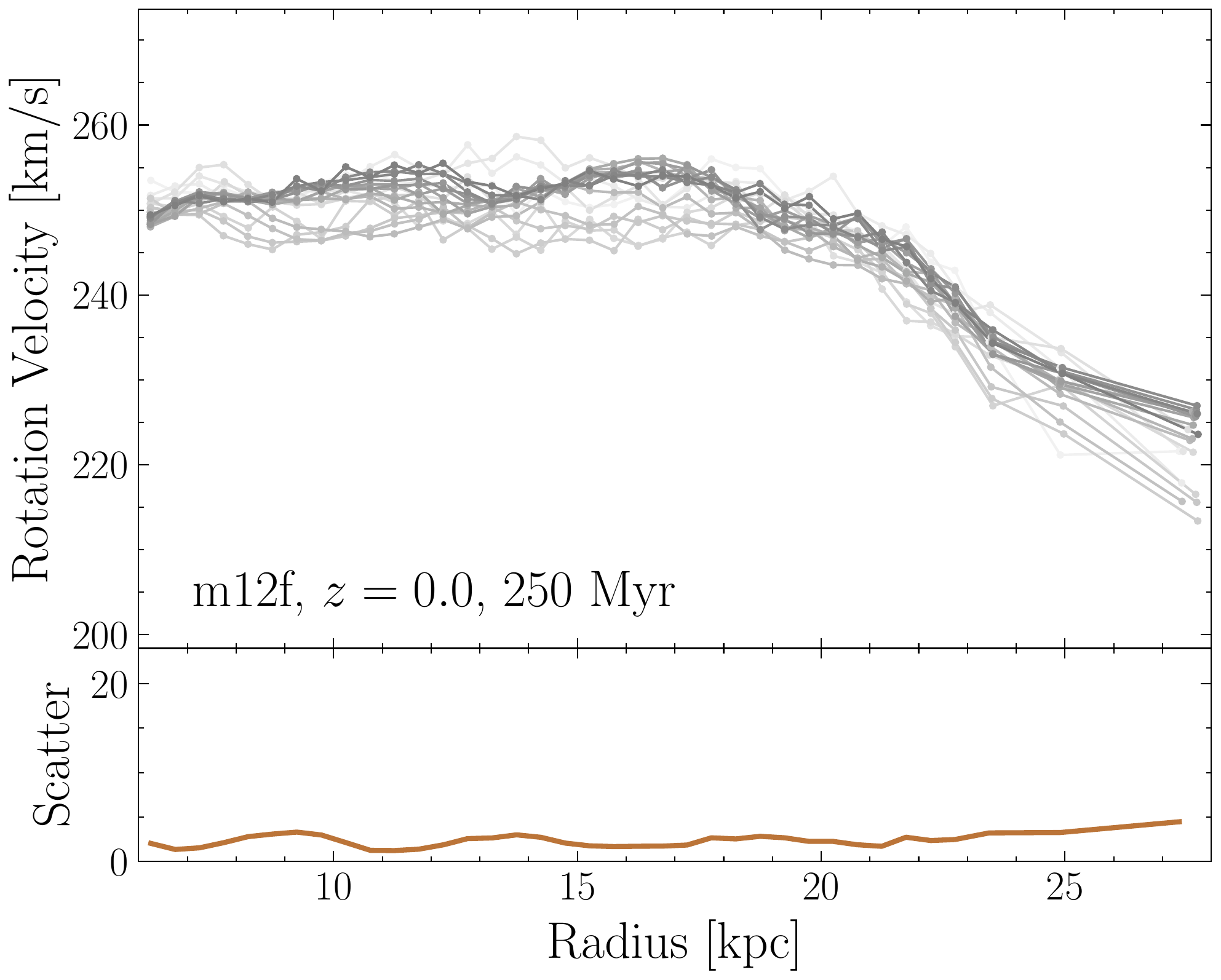} 
    \includegraphics[width=0.32\linewidth]{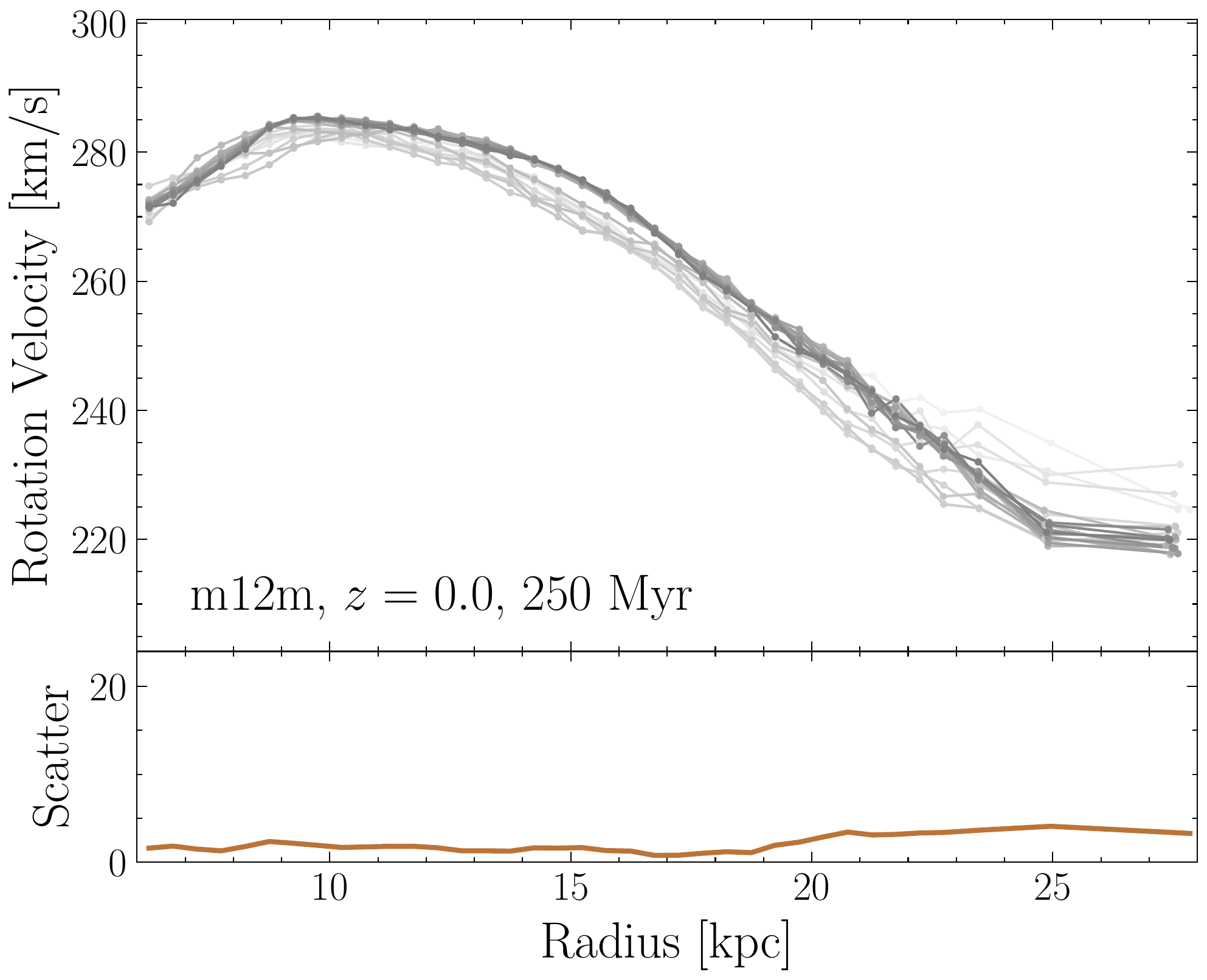}
    \caption{
    Comparison between the simulation snapshot curves (set (2)) within 250\,Myr from $z=0$.
    From left to right, the three plots show the results for different simulated galaxies at $z\sim0$ (snapshot 600): m12i, m12f, and m12m, respectively.
    For each galaxy, the top panel shows the derived rotation curve as a function of radius from the galactic center based on stars in snapshots from 580 to 600 ($z\sim0.02$-$0.0$).
    Similar to Figure~\ref{fig:sys_aa_rel}, the bottom panel shows the scatter in these curves. 
    The lighter shade of gray curves represent measured curves from earlier snapshots. 
    To mitigate existing axis asymmetry and highlight the time evolution, we average the curve across azimuthal angles for a given snapshot. 
    }
    \label{fig:sys_dd_rel}
\end{figure*}

The rotation curves derived from the simulation snapshot particles (set (2)) at different times (snapshots) following Jeans' equations show a moderate scatter when averaged over azimuthal angle in Figure~\ref{fig:sys_dd_rel}. 
For each snapshot, we first calculate the six rotation curves at different azimuthal angles as described in Section~\ref{sec:res_aa}.
We then take the average of these curves to find the azimuthal angle-averaged curve for each simulation snapshot.
Such a practice prevents the double-counting of uncertainties from the axis asymmetry described in Section~\ref{sec:res_aa}.
The different color lines represent the average rotation curves from snapshots within the typical dynamical timescales of thin disk stars: $\sim250$\,Myr.
We expect the scatter in the curves from consecutive redshifts to reflect the scale of dynamical disequilibrium.
The overall scatter size is $\lesssim10$\,\kmsec.

Another potential approach for avoiding doubling-counting is tracking and rotating the 60\,\degree wedges, equivalently the solar positions, across snapshots.
Such tracking is complicated because the star particles move with different velocities in $\phi$ and mix between wedges at earlier snapshots.
Thus, the $\phi=0$ reference points are arbitrarily set for all snapshots, except the $z=0$ to match the \Ananke\ reference frame, and we average curves over azimuthal angles at a given snapshot.

Thus, in comparison with the intrinsic scatter in the azimuthal angles, the scatter in the means of these measured curves is insignificant.
The only exception is a slightly greater scatter in m12i at large $R$, although still subdominant when compared with its scatter in azimuthal angles.
This low scatter in the mean suggests that, without a recent merger, we may safely assume dynamical equilibrium.

\begin{figure}
    \centering
    \includegraphics[width=0.95\linewidth]{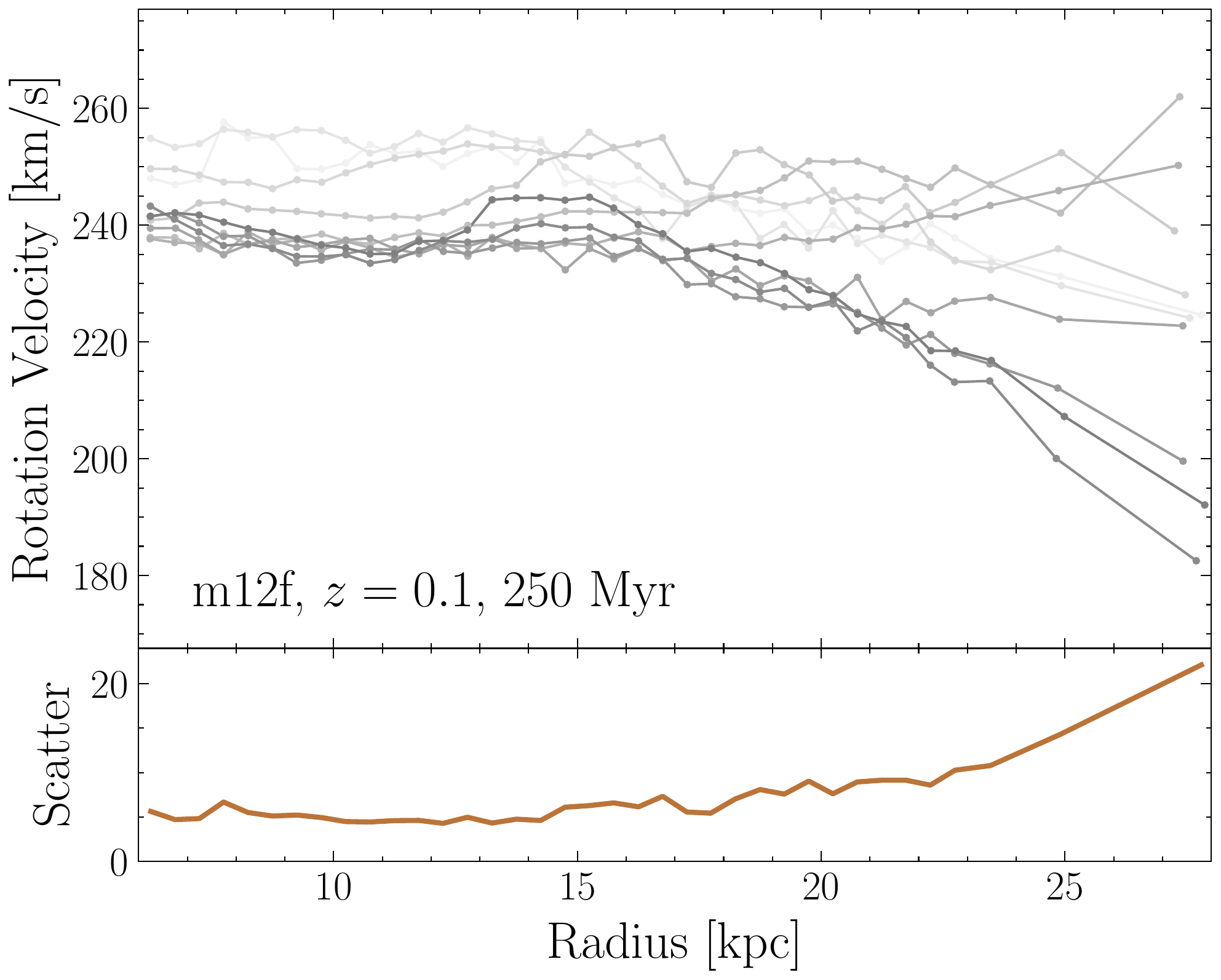}
    \caption{
    Comparison between the simulation snapshot curves (set (2)) within 250\,Myr from $z\sim0.1$ for m12f.
    The lighter shade of gray curves represent measured curves from earlier snapshots. 
    To mitigate existing axis-asymmetry and highlight the time evolution, we average the curve across azimuthal angles for a given snapshot. 
    The more significant scatter is driven by the merger event as described in Section~\ref{sec:res_dd}.
    }
    \label{fig:sys_dd_merger_rel}
\end{figure}

On the other hand, the rotation curves can show significant time evolution in the presence of recent mergers.
Specifically, given the typical dynamical timescale of $\sim250$\,Myr for disk stars, merger events with mass infalling on a similar timescale introduce time dependence to the gravitational potential.

We examine a known merger event at $z\sim0.1$ in m12f to quantify the effects of time-varying potential on the measured rotation curves.
The merger event was characterized by \citet{necib19b}, with stellar mass $\sim2\times10^9\,M_{\odot}$.
We find that the merger event can drive scattering in the rotation curve up to $\sim20$\,\kmsec.
We note that the effect of the merger events on the gravitational potential and calculated rotation curve is expected to depend on the infall time and orbital parameters.
For the Milky Way, the LMC and Sgr infall have been examined in the past for effects on the Milky Way dark matter halo \citep[e.g.,][]{vasiliev21,vasiliev23}.
Future studies with synthetic surveys of simulated Milky Way-like galaxies with both LMC and Sgr analogs \citep[e.g.,][]{buch24} are needed to quantify the exact impact of these merger events on the rotation curve measurements. 

\section{Discussion and Interpretation}
\label{sec:discussion}

\subsection{Non-negligible systematic biases and uncertainties}

We identify non-trivial systematic biases and uncertainties ($\gtrsim 5\%$) in all three simulated galaxies in the galactic radii range considered.
At larger galactic radii, generally, total uncertainties from various sources can amount to a significant $\sim 20\%$.
We summarize the four sources of systematics biases and uncertainties in Table~\ref{tab:sys_sum}.

\begin{deluxetable*}{lrrrrrrrr}
\caption{
Summary of systematic uncertainties from the four failure modes in this study for each of the three galaxies.
\label{tab:sys_sum}
}
\tablehead{
\colhead{} & \multicolumn{2}{c}{Biased population} & \multicolumn{2}{c}{Asym. drift correction} & \multicolumn{2}{c}{Axis asymmetry} & \multicolumn{2}{c}{Dynamical disequilibrium}\tablenotemark{$\ast$}  \\
\hline
\colhead{} & \colhead{Max. bias} & \colhead{Direction} & \colhead{Max. bias} & \colhead{Direction} & \colhead{Min. unc.} & \colhead{Max. unc.} & \colhead{Min. unc.} & \colhead{Max. unc.}
}
\startdata
m12i & $7\%$ & Positive\tablenotemark{$\dagger$} & $16\%$ & Negative & $1\%$ & $17\%$ & $0.4\%$ & $7\%$ \\
m12f & $16\%$ & Positive & $7\%$ & Negative & $2\%$ & $7\%$ & $0.5\%$ & $2\%$ \\
m12m & $6\%$ & Positive & $12\%$ & Negative & $0.4\%$ & $5\%$ & $0.3\%$ & $2\%$
\enddata
\tablenotetext{\ast}{Values reported here are estimated from the $z=0$ snapshots of the galaxies, where there are no recent major merger events. See Section~\ref{sec:res_dd} for a case of recent merger effects on m12f.}
\tablenotetext{\dagger}{The data point at the largest $R$ in Figure~\ref{fig:sys_sf_rel} for m12i has technically the greatest bias ($9\%$) in the negative direction. Since this is a single data point with a large statistical uncertainty and the rest of the data points all show positive bias, we opt to report the greatest bias in the positive direction here.}
\end{deluxetable*}

For the biases arising from the selection function and high-order asymmetric drift correction, the deviation tends to be in one direction: positive for the selection function and negative for the drift correction.
These preferential biases are expected due to how they are introduced in the first place, as discussed in Section~\ref{sec:results}, with magnitudes $10$-$40$\,\kmsec.
Furthermore, while these two biases are in opposite directions, the negative bias from the inaccurate drift correction is generally greater in magnitude than the positive bias from the selection function, as shown in Figure~\ref{fig:sys_sum_rel}.
The combined systematic biases from these two failure modes are thus preferentially biasing the rotation curves to be lower than the ground truth at the outer disk.
Ideally, one can quantify the extent of such directional biases and improve the Milky Way rotation curve calculation via either manual calibration or asymmetric uncertainties.
Doing so, however, requires a larger sample of simulated galaxies for statistics, and it is challenging to further quantify with our case study of only three galaxies.
We thus conservatively treat these two sources of systematic biases as symmetric when applying it to the Milky Way rotation curve in Section~\ref{sec:comp_mw}.

The axis asymmetry and dynamical disequilibrium introduce additional uncertainties in the measured curve, also on the order of $10$-$40$\,\kmsec.
There is no preferential direction for the biases from the axis asymmetry, as they are, by definition, the scatter in the curve from measuring the curve at different azimuth angles.
On the other hand, dynamical disequilibrium may have a preferred direction if caused by a recent merger event, as shown in the case of m12f at $z\sim0.1$.
For the Milky Way, recent simulations have indicated significant impacts on the Milky Way disk dynamics due to the ongoing infall of the LMC and Sgr \citep[see, e.g.,][]{vasiliev23,stelea24,asano25}.

\subsection{Galaxy-to-galaxy variance}

We observe variances in the dominating source of systematic uncertainties in the three simulated galaxies.
For m12i and m12m, the higher order correction dominates the total uncertainty, whereas the axis asymmetry and selection function effect dominates the uncertainty for m12f in the inner and outer disk, respectively, as shown in Figure~\ref{fig:sys_sum_rel}. 

In addition, for m12m, the other sources of systematic are consistently below 5\% for all galactic radii, while m12i experiences growth in other sources of systematic uncertainties towards the outer disk. 
We suspect this is due to the higher stellar mass and, thus, a larger stellar disk of m12m ($M_{*}\sim10^{11}\,M_{\odot}$) compared to m12i ($M_{*}\sim10^{10.8}\,M_{\odot}$) \citep{2023ApJS..265...44W}.
The quiet formation history of m12m may also contribute to the minimal contribution of systematic uncertainties from the other sources.

\subsection{Applying the new systematic uncertainties to the Milky Way rotation curve}
\label{sec:comp_mw}

We apply the total relative systematic uncertainties (shown in Figure~\ref{fig:sys_sum_rel}) derived from m12i in this study to the Milky Way rotation curve measurements from \citet{ou24}.
Since we aim for a conservative test, we adopt both the bias and uncertainty from the four failure modes as symmetric uncertainties from m12i. 
We selected m12i because it has the largest total systematic uncertainties among the three simulated galaxies, but note that the results remain qualitatively unchanged if we adopt the total systematic uncertainties from m12f or m12m. 
Because the m12i simulated galaxy and Milky Way do not necessarily have the same mass, we rescale the normalization of the uncertainties by multiplying the relative uncertainties from m12i by the measured velocities of the Milky Way.
We note that the total uncertainty on the rotation curve is dominated by the systematic uncertainty discussed in this work, as the statistical uncertainties are found to be negligible ($\sim0.5\%$).

We then redo the dark matter profile analysis as described \citet{ou24} and examine the goodness of fit using the generalized NFW (gNFW) and Einasto profile \citep{einasto65,navarro97}.
Details regarding the fit procedure and assumed baryonic components can be found in \citet{ou24} and references therein.
Briefly, for the gNFW profile, we adopt the density profile of the form
\begin{equation}
    \rho_{\rm{gNFW}} (r) = \frac{M_0}{4 \pi r_s^3} \frac{1}{(r/r_s)^\beta(1+r/r_s)^{3-\beta}},
    \label{eq:dens_gNFW}
\end{equation}
where $M_0$ is the mass normalization, $r_s$ is the scale radius, and $\beta$ is the characteristic power for the inner part of the potential. We recover the standard NFW profile for $\beta=1$. The Einasto profile is defined as
\begin{equation}
    \rho_{\rm{Ein}} (r) = \frac{M_0}{4 \pi r_s^3} \exp{\left(-(r/r_s)^\alpha\right)},
    \label{eq:dens_einasto}
\end{equation}
where $M_0$ and $r_s$ are defined similarly as in Equation~\ref{eq:dens_gNFW}, and $\alpha$ determines how fast the density distribution falls with galactic radius.
We additionally derive the virial mass ($M_{200}$), virial radius ($r_{200}$), concentration ($c_{200}$), the local dark matter density ($\rho_{\rm{DM},\odot}$) and the integrated $J$-factor at $15^\circ$ view angle.

\begin{figure*}
    \centering
    \includegraphics[width=0.45\textwidth]{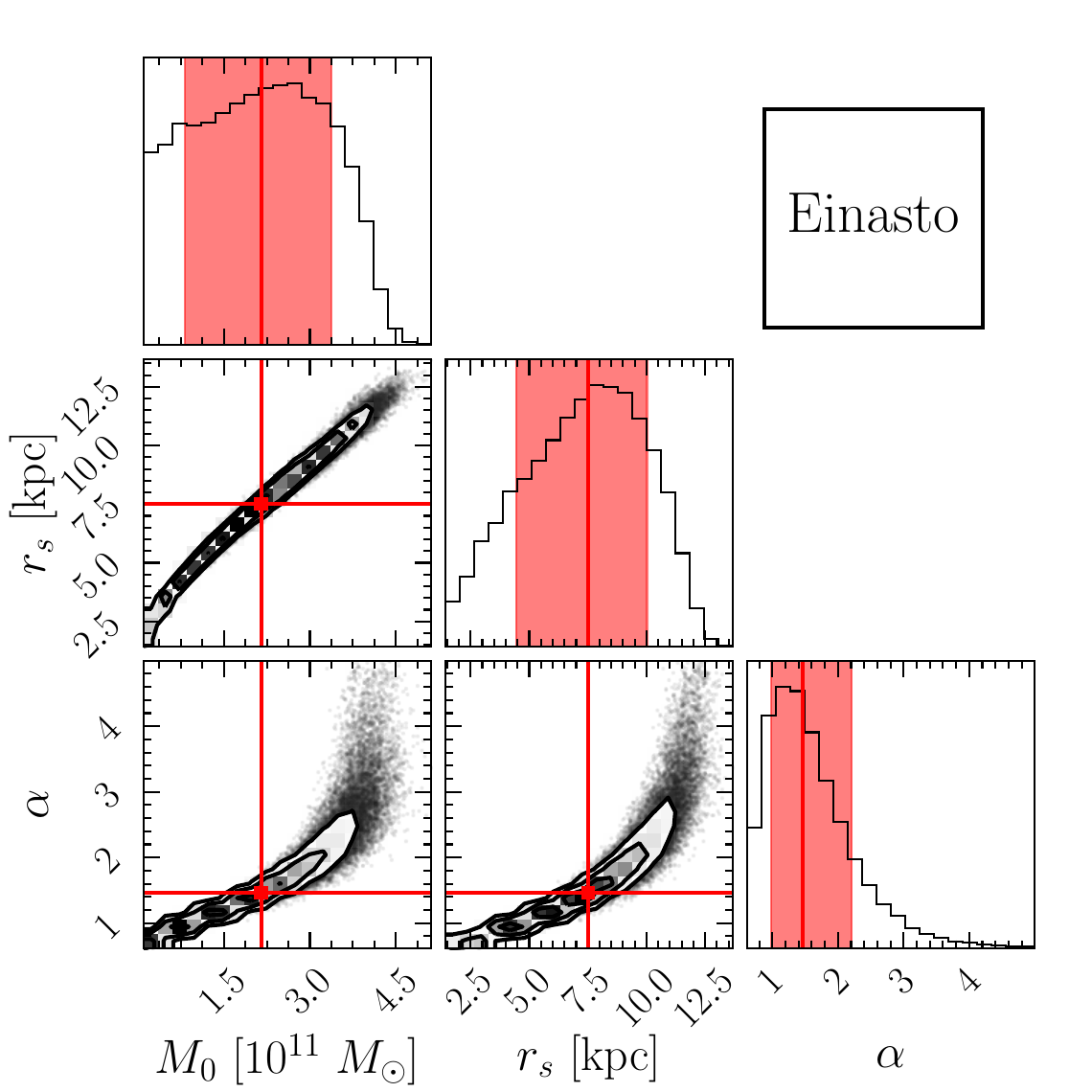}
    \includegraphics[width=0.45\textwidth]{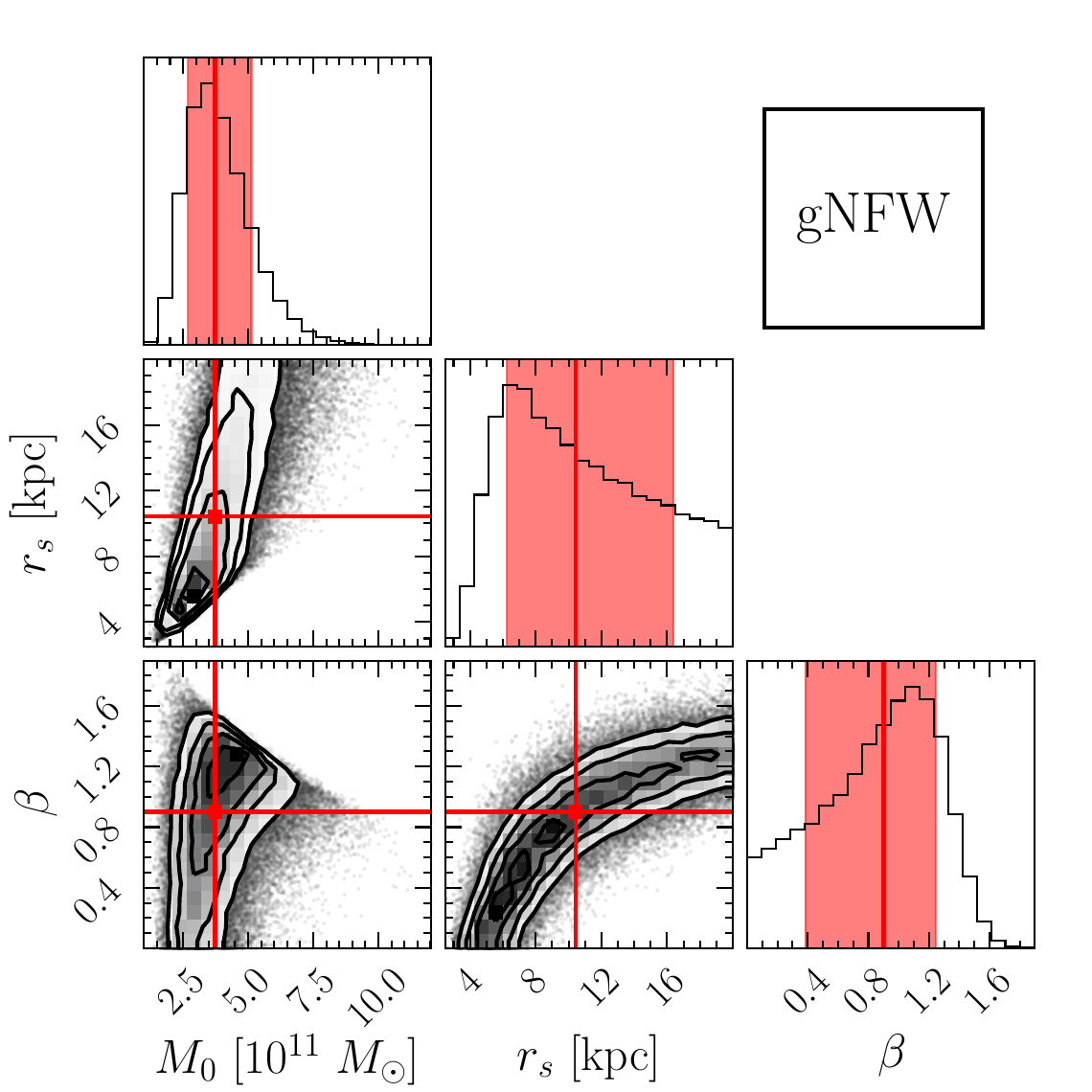}
    \caption{Posterior distribution of parameters for the Einasto (left) and gNFW (right) profile fit. The red line marks the median of the distribution, with the shaded region representing the 16$^{\rm{th}}$ to 84$^{\rm{th}}$ percentile. The gNFW inner slope parameter $\beta$ and scale radius $r$ posterior are non-Gaussian. }
    \label{fig:post_dtb}
\end{figure*}

The best-fit parameters and the derived quantities for both profiles are shown in Table~\ref{tab:fit_res}.
As expected, the increased uncertainties at all radii result in less constraining power for the profile parameters, as shown in Figure~\ref{fig:post_dtb}.
The posterior distribution has long extended tails that may be sensitive to the prior boundary, especially for the gNFW fit.
The posterior distribution for gNFW shows non-Gaussian features, with the distribution drifting towards a larger scale radius ($r_s$) and lower slope ($\beta$). 
The tail of the posterior distribution for $r_s$ is sensitive to the upper limit of the prior, with the posterior distribution tail extending with the prior.
This behavior suggests that the scale radius is poorly constrained with the increased uncertainty in the measurements. 
We limit the prior for $r_s$ to $[0,20]$\,kpc, as larger scale radii become less physical.
The Einasto fit also shows posterior distributions that are non-Gaussian, and in this case drift towards low mass and greater $\alpha$. 
Again, this is likely a result of the increased uncertainties.
We thus caution that the uncertainties of best-fit parameters should be treated as lower limits.

Compared with \citet{ou24}, the Einasto fit result is consistent despite the larger uncertainty. 
The Einasto profile is still cored with exponential decay in the density outside of $\sim10$\,kpc. 
Yet, the gNFW fit result is significantly different, especially in the inner slope $\beta$, which is now $\sim1$, consistent with the NFW profile.
This cannot be simply explained only by the overall increase in the uncertainties; it is also a result of the greater uncertainties at larger radii.
The greater uncertainties near the decline allow the gNFW fit to place less weight on the outer disk, preventing the decline from driving $\beta$ to $0$.
Meanwhile, the Einasto profile fit is flexible enough by design that the parameters are still driven to a cored profile despite the large uncertainties at the outer disk.

\begin{deluxetable*}{lccc}
\caption{\textsc{emcee} fitted results for Einasto and gNFW dark matter halo profiles. We report the median of the posterior distribution and the 16$^{\rm{th}}$ to 84$^{\rm{th}}$ percentile as uncertainties. 
NOTE: both profiles shown problematic behaviors in the posterior distribution and are thus only shown here for completeness as discussed in Section~\ref{sec:comp_mw}. 
The virial masses and radii are inconsistent at $2\,\sigma$ between the two best-fit profiles, suggesting the physical interpretation of the same data is sensitive to the chosen functional form of the underlying dark matter profile. 
}
\label{tab:fit_res}
\tablehead{
\colhead{} & 
\colhead{Einasto} & 
\colhead{gNFW} & 
\colhead{Prior} 
}
\startdata
\textbf{Normalization Mass ($M_0$)} & \MnorE & \Mnorg & $[0.1,1000]\times10^{11}$~\msun\tablenotemark{$\ast$} \\
\textbf{Scale Radius ($r_s$)} & \rscaE & \rscag & $[0,20]$~kpc \\
\textbf{Slope Parameter ($\alpha,\beta$)} & \slopE & \slopg & $[0,5]$ \\
\hline 
\textbf{Virial Mass ($M_{200}$)} & \MvirE & \Mvirg & -- \\
\textbf{Virial Radius ($r_{200}$)} & \rvirE & \rvirg & -- \\
\textbf{Concentration ($c_{200}$)} & \concE & \concg & -- \\
\textbf{Local Dark Matter Density ($\rho_{\rm{DM},\odot}$)} & \rhoSE & \rhoSg & -- \\
\hline 
\textbf{$\chi^2$ per d.o.f. ($\chi_{\nu}^2$)} & \chisE & \chisg & --
\enddata
\tablenotetext{\ast}{Fitted in logarithmic scale.}
\end{deluxetable*}

\begin{figure*}
    \centering
    \includegraphics[width=\textwidth]{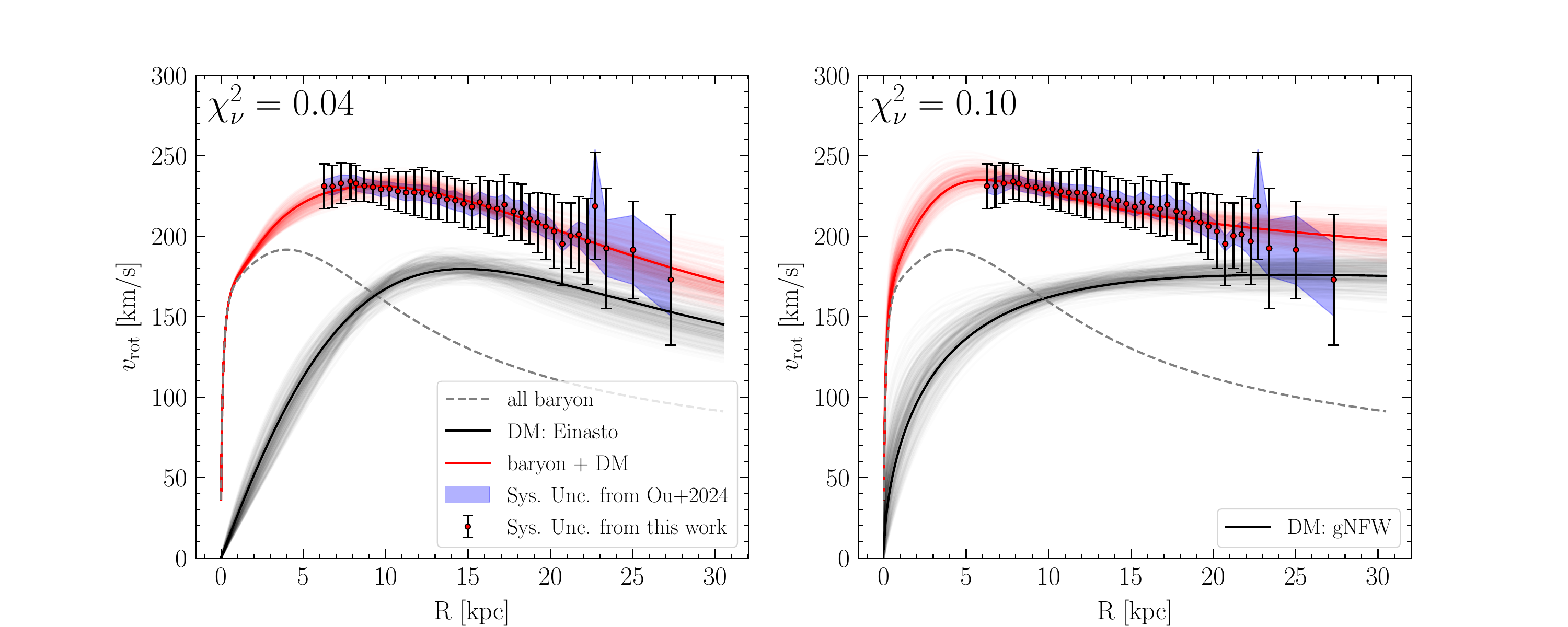}
    \caption{Comparison between Einasto(left) and gNFW(right) profile fit to the measurements from \citet{ou24} (red dots) with systematic uncertainties derived from m12i in this study (black error bars).
    Systematic uncertainties from \citet{ou24} (blue shaded region) is included for comparison.
    }
    \label{fig:halo_fit_compare}
\end{figure*}

Despite the broad posterior, the goodness of the fit, evaluated by the reduced $\chi^2$ values, is reasonable for both profiles, as shown in Figure~\ref{fig:halo_fit_compare}.
While the Einasto fit still has a lower $\chi^2$ value, the Bayes factor is only $\sim 2.5$ between the Einasto and gNFW profile fit due to increased uncertainties.
There is, thus, no statistically significant preference for the Einasto profile.

While the goodness of fit is comparable, we highlight the significantly different behaviors of the two profiles when extrapolated to inner and outer galactic radii. 
As shown in Figure~\ref{fig:menc}, the density profile towards the galactic center ($r\lesssim1$\,kpc) and the enclosed mass profile towards the outer halo ($r\gtrsim30$\,kpc) from the two profiles disagree at $>2\,\sigma$.
Such disagreement between the Einasto and gNFW results indicate that the physical interpretation of the same data is sensitive to the assumed functional form of the dark matter profile.
Thus, even with a proper treatment of the systematic uncertainties on the rotation curve measurements, mapping from the curve to the underlying dark matter profile is still not trivial.
We emphasize again that caution should be exercised when extrapolating the rotation curve measurements with fixed functional form assumptions.
With that in mind, we proceed to compare the best fit profiles with other literature constraints of the Milky Way mass distribution as independent external guide for examining the accuracy of the two best-fit profiles.

\subsubsection{Comparison with literature}
\label{sec:lit_comp}

\begin{figure*}
    \centering
    \includegraphics[width=\textwidth]{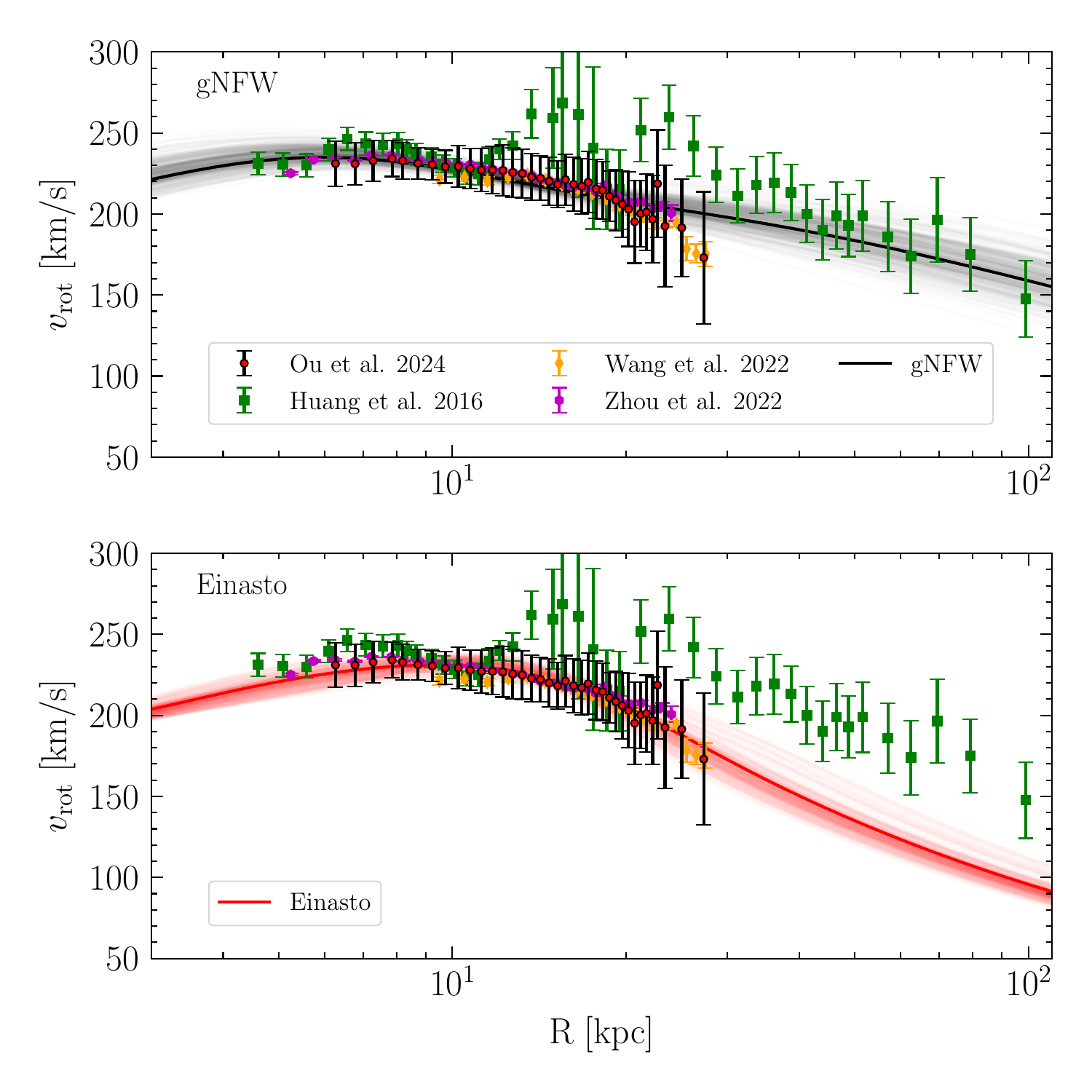}
    \caption{Best-fit model rotation curves using the gNFW (top black lines) and Einasto (bottom red lines) profiles fitted on measurements from \cite{ou24} (red dots) with systematic uncertainties derived from m12i in this study. Literature values for the rotation curve are taken directly from \citet{huang16}, \citet{wang22}, and \citet{zhou22}. 
    We emphasize that we only fit the measurements from \citet{ou24} (red dots) with the added systematic uncertainties from m12i in this study.
    Compared between the two panels, the extrapolated model rotation curves from gNFW (black lines) are more consistent with the rotation curve measurements from other probes at $R>30$\,kpc.
    }
    \label{fig:gnfw_einasto_w_lit}
\end{figure*}

We compare the new best fit profiles with the literature constraints, both from rotation curves and direct enclosed mass measurements.
Figure~\ref{fig:gnfw_einasto_w_lit} shows the gNFW and Einasto best fit rotation curves and the data compared with the rotation velocities from the literature \citep{huang16,wang22,zhou22}.
Despite being fitted only to the data from \citet{ou24}, the extrapolated model rotation curves from gNFW are in better agreement with the rotation curve measurements from \citet{huang16} at $R>30$\,kpc.

\begin{figure}
    \centering
    \includegraphics[width=\linewidth]{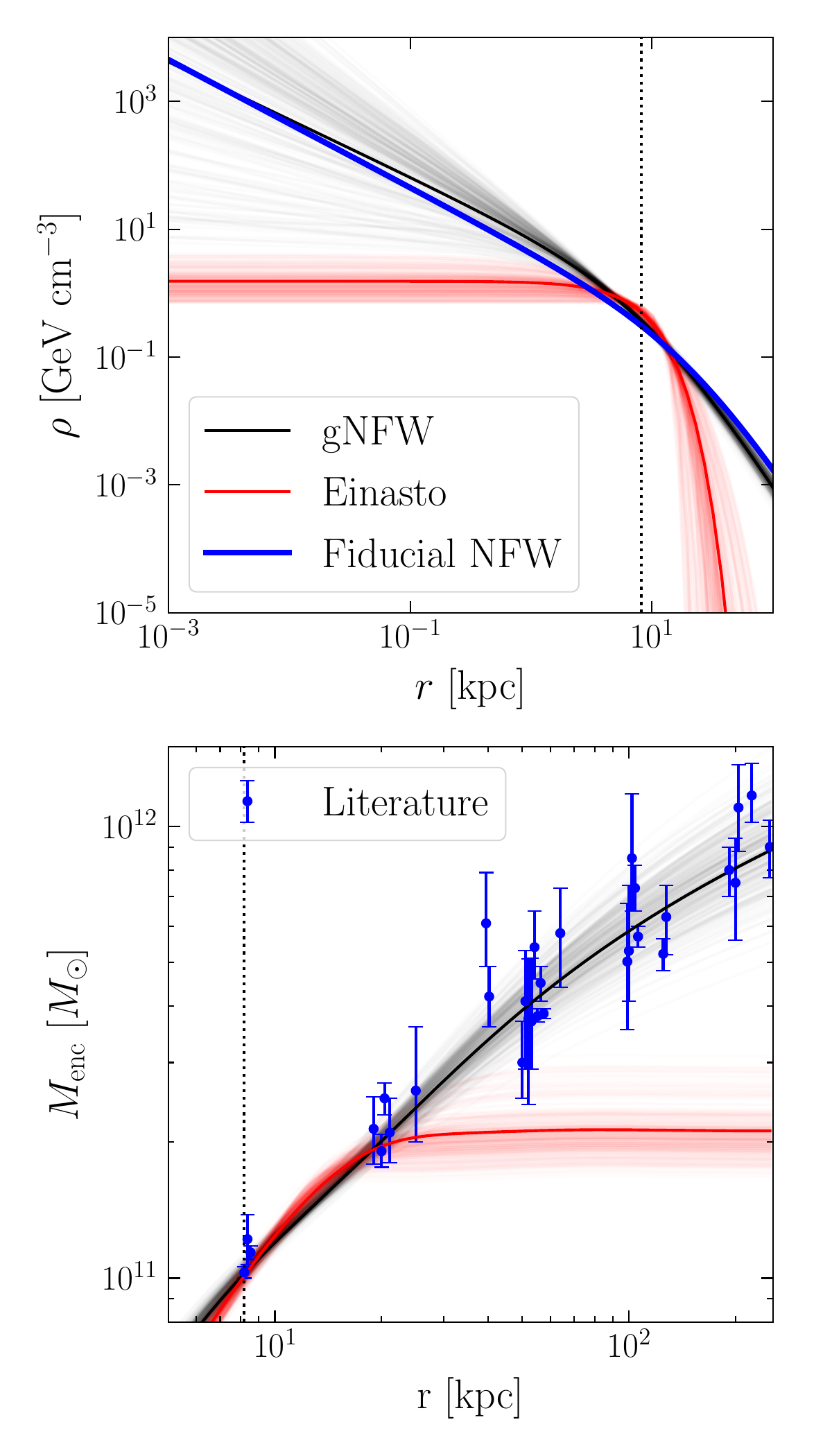}
    \caption{
    Density (top) and enclosed mass (bottom) profiles from the Einasto (red lines) and gNFW (black lines) posterior distributions. 
    The dotted lines in both panels mark the solar distance from the galactic center.
    The top panel shows that the posterior distribution of the gNFW density profiles are consistent with the fiducial NFW profile (blue lines), with a cuspy central density.
    On the other hand, the Einasto density profiles are cored and differ significantly from the gNFW at the inner Galaxy.
    The bottom panel shows literature constraints (gray dots) (see \citealt{wang20} and references therein) of the enclosed mass from globular clusters, dwarf galaxies, and stellar streams.
    The gNFW profiles are qualitatively consistent with the literature, while the Einasto profiles flatten out at $r>30$\,kpc, suggesting strong tension with results from other probes.
    }
    \label{fig:menc}
\end{figure}

When examined in terms of the enclosed mass, the gNFW best-fit profile is more consistent with the measurements from other probes that constrain the enclosed mass of the Milky Way at $R>30$\,kpc, as shown in Figure~\ref{fig:menc}.
Specifically, the Einasto best-fit profile disagrees with the enclosed mass estimate from the outer halo ($R\gtrsim100$\,kpc) at $>2\,\sigma$ level, while the gNFW best-fit profile agree well within 1\,$\sigma$ in most cases.
Thus, despite the comparable goodness of fit, the gNFW fit is preferable when factoring in the constraints from other studies at larger galactic radii.

In other words, assuming an underlying gNFW profile and incorporating the systematic uncertainties identified in this study, the extrapolated Milky Way rotation curve from \citet{ou24} is \textit{no longer} in tension with the Milky Way enclosed mass constraints from globular clusters and dwarf galaxies.

\section{Summary}

In this study, we examined the robustness of the Jeans procedure for deriving the rotation curve of the Milky Way and its implications for understanding the dark matter distribution. 
Using three Milky Way-mass galaxies from the \Fire-2 simulations and synthetic \Gaia DR3 surveys, we quantified the impact of systematic uncertainties introduced by various sources, including survey selection function biases, incorrect asymmetric drift corrections, dynamical disequilibrium from merger events, and axis asymmetry in the stellar population.

The key takeaways from this study are:
\begin{itemize}
    \item Our results reveal that systematic uncertainties can be significant, especially at large galactocentric distances, reaching up to 20\% in some cases. 

    \item We identify inaccurate asymmetric drift correction terms in the Jeans equation and axis asymmetry as the dominant contributors to these uncertainties in galaxies with no recent merger events.

    \item In the case of a recent merger event, the dynamical disequilibrium dominates the systematic uncertainties as expected.

    \item We show that the total systematic uncertainty in current Milky Way rotation curve measurements has been underestimated, which may explain the observed tension in dark matter halo mass estimates derived from different probes.
    We thus apply the systematic uncertainties identified in our simulations to the Milky Way rotation curve measurements and find that the large uncertainties weaken the evidence for a significant dark matter halo mass drop-off. When adopting the gNFW density profile to fit the rotation curve with the systematic uncertainty from this study, we find consistent results between the extrapolated best-fit profile and the enclosed mass profile from other probes.

\end{itemize}

Our analysis highlights the need for careful consideration of systematic biases in future measurements to better constrain the Milky Way's dark matter profile. More specifically, this study has shown that, going forward, a number of considerations should be taken into account when establishing the rotation of the Milky Way based on stars:

\begin{enumerate}
    \item Future rotation curve measurements for the Milky Way should account for the systematic biases and uncertainty estimates based on any assumptions made.

    \item Future analyses of the Milky Way potential should combine rotation curve measurements from various probes (e.g., from stellar tracers, dwarf galaxies, and stellar streams) to avoid producing results that are systematically biased. To avoid systematic biases. Appropriate care needs to be taken with the functional form used in any fit to observed data. 
\end{enumerate}

More broadly, ultimately, rotation curve measurements based on stellar tracers are systematics-limited -- even in the \Gaia\ era of high-precision astrometry. Unfortunately, these systematics are not a single-point failure that can easily be addressed yet they clearly need to be taken into account and explored further. We thus caution that 
\textit{measuring the rotation curve based on Galactic stars alone is no longer sufficient for studying the dark matter distribution in the Milky Way.}

Instead, we propose developing data-driven methodologies that directly map the potential to the observed phase-space distribution of the stars. Such an approach would allow for maximizing the information that can be recovered from stellar tracers. These new methodologies are independently testable with 
synthetic surveys (mock observations of stars in simulations) and cosmological galaxy formation simulations of Milky Way-like galaxies and can thus serve as important test grounds before applying them to existing and new observational data.

\begin{acknowledgements}
We thank Anna-Christina~Eilers for valuable discussions. 
X. O. thanks the LSST Discovery Alliance Data Science Fellowship Program, which is funded by LSST Discovery Alliance, NSF Cybertraining Grant \#1829740, the Brinson Foundation, and the Moore Foundation; his participation in the program has benefited this work.
XO is supported by NSF-AAG grant AST-2307436. 
LN is supported by the Sloan Fellowship, the NSF CAREER award 2337864, NSF award 2307788, and by the NSF award PHY2019786 (The NSF AI Institute for Artificial Intelligence and Fundamental Interactions, http://iaifi.org/).
AW received support from NSF, via CAREER award AST-2045928 and grant AST-2107772.

This work used Stampede-2 under allocation number TG-PHY210118, part of the Extreme Science and Engineering Discovery Environment (XSEDE), which is supported by National Science Foundation grant number ACI-1548562. This work used the Engaging cluster supported by the Massachusetts Institute of Technology.

This research has made use of NASA's Astrophysics Data System Bibliographic Services; the arXiv pre-print server operated by Cornell University; the SIMBAD and VizieR databases hosted by the Strasbourg Astronomical Data Center.

\end{acknowledgements}

\software{%
matplotlib \citep{hunter07},
numpy \citep{vanderwalt11},
scipy \citep{jones01},
emcee \citep{foremanmackey13},
astropy \citep{astropy:2013,astropy:2018}, and
galpy \citep{bovy15}.}

\clearpage

\bibliographystyle{aasjournal}
\bibliography{xou}



\appendix

\renewcommand{\thefigure}{A\arabic{figure}} 
\setcounter{figure}{0} 

\section{Potential axis asymmetry comparison for all galaxies}
\label{sec:more_pot}

We present the comparisons between the axis asymmetric fluctuation in potential and the stellar mass distribution in the galactic plane for m12i and m12m.

\begin{figure}[h]
    \centering
    \includegraphics[width=0.45\linewidth]{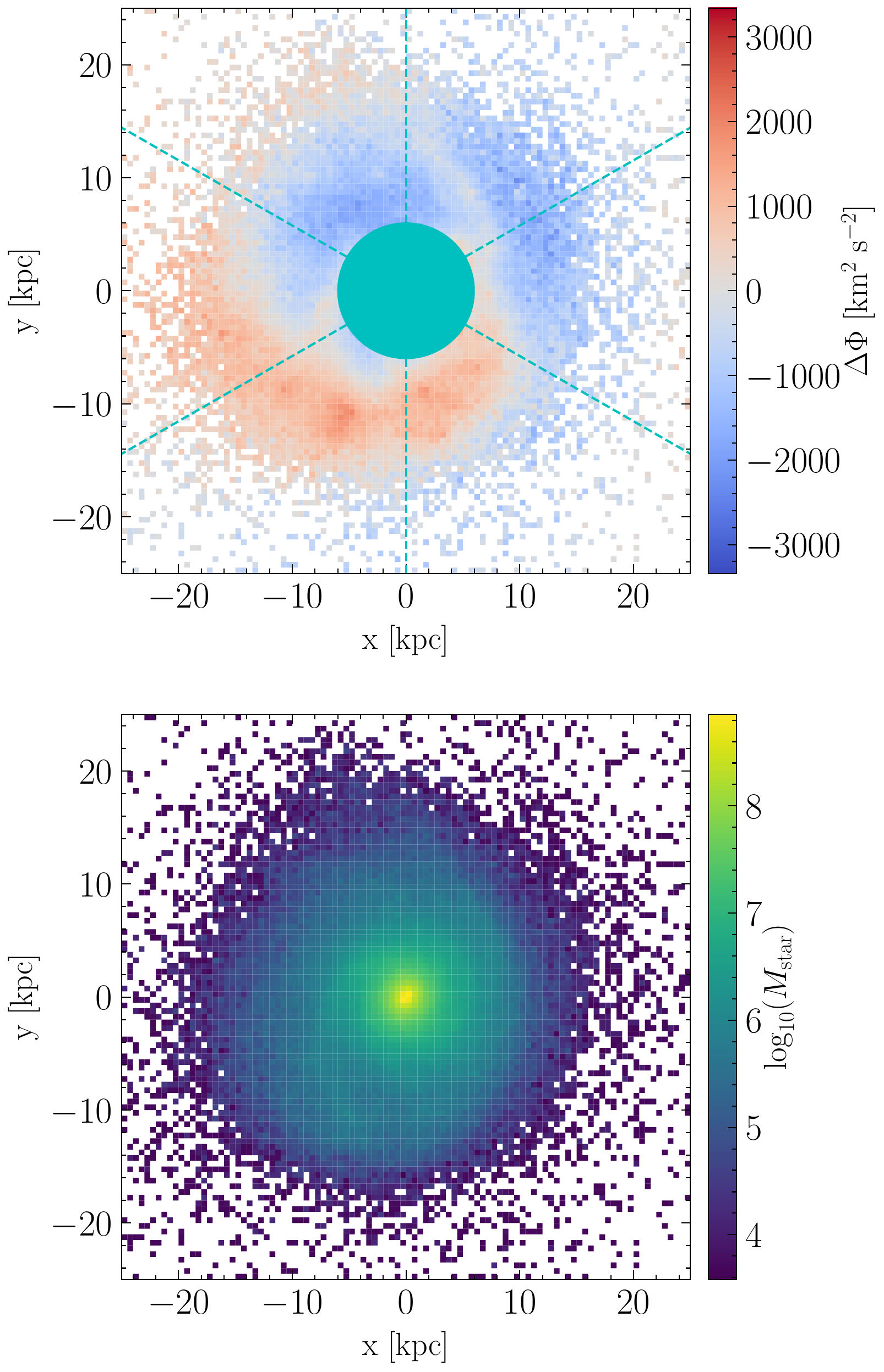} %
    \includegraphics[width=0.45\linewidth]{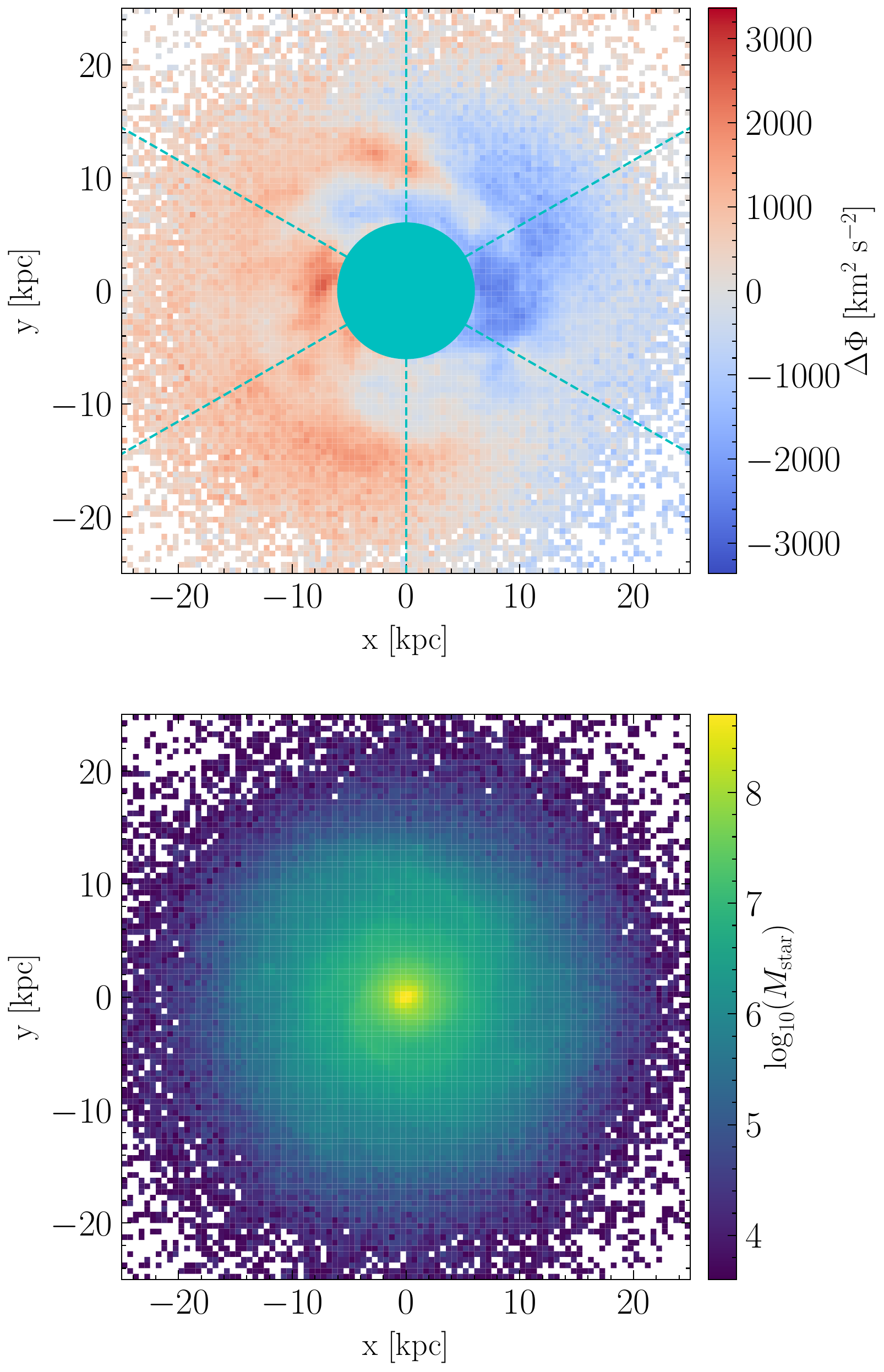} %
    \caption{
    Comparison between the axis asymmetric potential experienced by star particles (top) and the stellar mass (bottom) in the galactic plane for m12i (left) and m12m (right).
    The \agama-fitted axisymmetric component is subtracted to highlight the azimuthal variation in potential.
    The cyan mask at the center of the top panel indicates the region not employed in the rotation curve calculation due to known non-axisymmetric potential driven by the galactic bar.
    The dashed lines mark the different azimuthal slices used to compute the rotation curves.
    The deviation in gravitational potential from complete axis symmetry in the top panel matches qualitatively with the stellar mass distribution in the stellar disk.
    }
    \label{fig:delta_pot_map_extra}
\end{figure}

\end{CJK*}
\end{document}